\documentclass[aps,pra,10pt,notitlepage,nofootinbib,tightenlines,twocolumn,superscriptaddress]{revtex4-1}
\usepackage[utf8]{inputenc}

\usepackage{bm,bbm,amssymb,amsmath,amsfonts,amsthm,mathrsfs,MnSymbol, times}
\usepackage[toc,page]{appendix}
\usepackage{natbib}
\usepackage{graphicx}
\usepackage{multirow}

\usepackage{color}
\usepackage{bbold}
\usepackage[utf8]{inputenc}
\usepackage{empheq}
\usepackage{soul}
\usepackage[ampersand]{easylist} 
\usepackage[normalem]{ulem}
\usepackage{braket}
\usepackage{hyperref}
\hypersetup{
    colorlinks=true,       
    linkcolor=red,          
    citecolor=magenta,        
    filecolor=magenta,      
    urlcolor=cyan,           
    runcolor=cyan
}
\graphicspath{ {figures/} }

\def\sx{\sigma^x}
\def\sy{\sigma^y}
\def\sz{\sigma^z}

\def\Tr{\text{Tr}}
\def\xy4{{\rm XY4}}
\def\xyp{{\rm XY4}${}^\prime$}
\def\crxy4{{\rm CR}-{\rm XY4}}

\DeclareMathOperator{\sgn}{sgn}

\newcommand{\eq}[1]{\begin{align}#1\end{align}}
\newcommand{\mrm}{\mathrm}

\newcommand{\mM}{\mathcal{M}}
\newcommand{\floor}[1]{\lfloor #1 \rfloor}
\newcommand{\ffr}{\floor{\frac{r}{2}}}

\def\xy4{{\rm XY4}}
\def\xyp{{\rm XY4}${}^\prime$}
\def\crxy4{{\rm CR}-{\rm XY4}}

\begin{document}
\title{Quantum Crosstalk Robust Quantum Control}
\author{Zeyuan Zhou}
\affiliation{William H. Miller III Department of
Physics $\&$ Astronomy, Johns Hopkins University, Baltimore, Maryland 21218, USA}
\author{Ryan Sitler}
\affiliation{Johns Hopkins University Applied Physics Laboratory}  
\author{Yasuo Oda}
\affiliation{William H. Miller III Department of
Physics $\&$ Astronomy, Johns Hopkins University, Baltimore, Maryland 21218, USA}
\author{Kevin Schultz}
\affiliation{Johns Hopkins University Applied Physics Laboratory}  
\author{Gregory Quiroz}
\affiliation{Johns Hopkins University Applied Physics Laboratory}       
\affiliation{William H. Miller III Department of
Physics $\&$ Astronomy, Johns Hopkins University, Baltimore, Maryland 21218, USA}

\begin{abstract}
    The prevalence of quantum crosstalk in current quantum devices poses challenges for achieving high-fidelity quantum logic operations and reliable quantum processing. Through quantum control theory, we develop an analytical condition for achieving crosstalk-robust single-qubit control of multi-qubit systems. We examine the effects of quantum crosstalk via a cumulant expansion and develop a condition to suppress the leading order contributions to the dynamics. The efficacy of the condition is illustrated in the domains of quantum state preservation and noise characterization through the development of crosstalk-robust dynamical decoupling and quantum noise spectroscopy (QNS) protocols. Using the IBM Quantum Experience, crosstalk-robust state preservation is demonstrated on 27 qubits, where a $3\times$
    improvement in coherence decay is observed for single-qubit product and multipartite entangled states. Through the use of noise injection, we experimentally demonstrate crosstalk-robust dephasing QNS on a seven qubit processor, where a $10^4$ improvement in reconstruction accuracy over ``cross-susceptible" alternatives is found. Together, these experiments highlight the significant impact the crosstalk mitigation condition can have on improving multi-qubit characterization and control on current quantum devices.
\end{abstract}
\maketitle

%
%
The ability to implement high-fidelity quantum logic operations is a necessity for achieving reliable and scalable quantum computing~\cite{Preskill2018quantum}. In current systems, however, system-environment interaction and crosstalk are typically quite substantial and ultimately limit qubit coherence and gate accuracies. Characterized as unwanted inter-qubit coupling, quantum crosstalk causes undesired dynamics that violate the locality and individual addressability of qubits. It is known to be prevalent in current systems and present obstacles in the implementation of quantum algorithms~\cite{Ding2020compilation,Murali2020scheduler,Babukhin2021noise,ohkura2022parallelqc,Xie2022optimization}, quantum characterization~\cite{Gambetta2012benchmarking,mckay2019rb3q,sarovar2020xtalk,rudinger2021gst}, and instantiations of quantum error correction~\cite{Chen2021qec,Parrado-Rodriguez2021trappedion}.

Quantum crosstalk arises in a variety of qubit architectures. Atomic systems are susceptible to unwanted interactions between neighboring spectator qubits during two-qubit operations~\cite{ospelkaus2008trapped, urban2009rydberg, Auger2017Rydberg, Levine2018Rydberg, Parrado-Rodriguez2021trappedion,Fang2022trapped-ion}, while superconductor (e.g., fixed-frequency transmons~\cite{krantz2019quanteng, Ash-Saki2020crosstalk,kandala2021transmons,zhao2022transmon}) and semiconductor~\cite{Buterakos2018semiconductor, Heinz2022spinqubit,throckmorton2022spinqubit} platforms commonly experience parasitic $ZZ$ interactions from always-on coupling used to implement entangling gates. Various strategies have been proposed to address crosstalk from the hardware and software perspective. Hardware solutions have predominately centered around architecture design~\cite{Mundada2019couplertuning,Ku2020hybrid,Zhao2020anharmonicity}. Software approaches are diverse and have sought to address crosstalk at the physical~\cite{Buterakos2018semiconductor,carvalho2021xtalk,Wei2021AC-Stark,Vinay2021DD,Fang2022trapped-ion,Xie2022optimization} and compiler~\cite{Murali2020scheduler,Ding2020compilation,zhang2020slackq,Zhang2022inverse} layers of the quantum software stack. Despite their utility, these approaches are either hardware-specific or provide limited insight and intuition into broader principles for crosstalk mitigation.

We address this challenge by leveraging quantum control theory to develop an analytical condition for achieving crosstalk robust single-qubit control of multi-qubit systems. Quantum control is a widely used tool for constructing high-fidelity gates~\cite{dong2010qctrlreview, Green2013control,ball2015walsh,wilhelm2020qctrl, dAlessandro2021ctrlbook,Oda2022grafs} and error mitigation strategies~\cite{viola1999dd,biercuk2011dd,lidar2013qec-book,Pokharel2018DD}, as well as unraveling key characteristics of spatio-temporally correlated noise through quantum noise spectroscopy (QNS)~\cite{alvarez2011qns,bylander2011qns,cywinski2014qns,norris2016nongauss, Paz-Silva2017QNS,frey2017slepctrl,sung2019qns,frey2020slepxyz,vonlupke2020twoqubitqns,Lawrie2021semiconductor}. We exploit a control framework commonly used to derive the filter function formalism~\cite{cywinski2008fff,Green2013control,ball2015walsh,Paz-Silva2017QNS,norris2018slepian,Chalermpusitarak2021fff} to examine the impact of crosstalk on system dynamics. Through the use of a perturbative cumulant-based expansion, we derive a control condition that enables quantum crosstalk cancellation up to second order in the total evolution time.

The versatility and applicability of our approach is illustrated through experimental investigations of crosstalk robust quantum state preservation and noise characterization on the IBM Quantum Experience (IBMQE). Crosstalk-robust dynamical decoupling (CRDD) is introduced and shown to dramatically improve the simultaneous preservation of single-qubit product states (SPSs) and multipartite entangled states (MESs) up to 27 qubits. Furthermore, we demonstrate the relevance of the condition to noise characterization, where a crosstalk-robust dephasing QNS protocol is introduced and subsequently used to perform the first known simultaneous noise spectrum estimation on seven qubits. Together, these experiments highlight the deleterious effects of crosstalk and the remarkable impact our condition can have on improving multi-qubit characterization and control on current quantum devices.

%
%
\emph{Crosstalk Noise Model.} -- We focus on the suppression of crosstalk during the implementation of single qubit operations as this represents the most fundamental type of control one may possess on a quantum system. To this end, we consider an $N$ qubit system subject to noisy, controlled evolution governed by the time-dependent Hamiltonian $H(t) = H_C(t) + H_E(t)$. The control Hamiltonian is given by 
\begin{equation}
    H_{C}(t) = \sum\limits_{i=1}^{N} \frac{\Omega_i(t)}{2}(\sigma_{i}^{x}\cos\phi_{i}(t) + \sigma_{i}^{y}\sin\phi_{i}(t)),
\end{equation}
where $\Omega_i(t)$ and $\phi_i(t)$ represent the time-dependent control amplitude and phase, respectively. Noise is generated by the error Hamiltonian 
\begin{equation}
    H_E(t) = \sum\limits_{i=1}^{N}\vec{\sigma_{i}}\cdot\vec{\beta_i}(t) + \sum^N_{i< j} J_{ij}\,\sigma_i^{z}\sigma_j^{z},
    \label{eq:H-err}
\end{equation}
with $\vec{\sigma}_i=(\sx_i,\sy_i,\sz_i)$ consisting of the $i$th qubit Pauli operators. All relevant noise contributions are captured by semi-classical, spatio-temporally correlated noise processes and static quantum crosstalk. The former is described by
$\vec{\beta_i}(t) = (\beta_i^x(t),\beta_i^y(t),\beta_i^z(t))$, where $\beta^\mu_i(t)$ is assumed to be a wide-sense stationary Gaussian
stochastic process with zero-mean, $\overline{\beta_i^\mu(t)}=0$, and two-point correlation functions $C^{\mu\nu}_{ij}(\tau)=\overline{\beta^\mu_i(\tau)\beta^\nu_j(0)}$, $\mu,\nu=x,y,z$. Note that $\overline{\cdots}$ denotes classical ensemble averaging. The crosstalk is characterized by the static coupling strength $J_{ij}$ and 2-local $ZZ$ interactions. This model and its generalization~\cite{supp} 
are relevant to a wide range of experimental platforms, including superconducting qubits~\cite{bylander2011qns,krantz2019quanteng, Ash-Saki2020crosstalk, Ku2020hybrid}, semiconductor qubits~\cite{Buterakos2018semiconductor,Lawrie2021semiconductor,connors2022qns,Heinz2022spinqubit}, and trapped ion systems~\cite{ospelkaus2008trapped, frey2017slepctrl,frey2020slepxyz,he2021qns, Parrado-Rodriguez2021trappedion,Fang2022trapped-ion}.

%
%
\emph{Effective Error Dynamics.} -- We investigate the effect of $H_E(t)$ on the dynamics of expectation values and fidelity using time-dependent perturbation theory. In the limit of strong control and weak noise~\cite{Paz-Silva2017QNS,quiroz2021qaoa,Chalermpusitarak2021fff}, the dynamics are dominated by the control such that $H_E(t)$ is treated as a perturbation. The evolution generated by the error Hamiltonian is isolated by moving into the interaction (toggling) frame with respect to the control such that the total evolution is described by $U(T)=\tilde{U}_E(T)U_C(T)$. The propagator $U_C(T)=U_C(T,0)=\mathcal{T}_+e^{-i\int^{T}_0dt\,H_C(t)}$ describes the control dynamics, while $\tilde{U}_E(T)=\mathcal{T}_+e^{-i\int^{T}_0dt\,\tilde{H}_E(t)}$ is the rotated-frame time evolution governed by $\tilde{H}_E(t)=U_C(T,t)H_E(t)U_C^\dagger(T,t)$; $\mathcal{T}_{+}$ denotes the time-ordering operator. The rotated-frame error Hamiltonian is further specified by
\begin{equation}
    \tilde{H}_E(t) = \sum_{i=1}^N \vec{\Lambda}_i(t) \cdot \vec{\beta}_i(t)  + \sum_{i,j=1}^N J_{ij} [\vec{\Lambda}^T_i(t)]_z[\vec{\Lambda}_j(t)]_z,
\end{equation}
where $\vec{\Lambda}_i(t)\equiv\boldsymbol{R}_i(t) \vec{\sigma}^T_i$ and $\boldsymbol{R}_i(t)$ is the  ``control matrix" with elements ${R}^{\mu\nu}_i(t)=\Tr \left[ U_{C}(T,t)\sigma_{i}^\mu U^{\dagger}_{C}(T,t)\sigma_{i}^\nu \right]/2$. $A^T$ and $[\vec{a}]_z$ denote the transpose of $A$ and the $z$-component of $\vec{a}$, respectively.

We quantify the impact of noise on the time-dependent expectation value of an observable $O$ using a cumulant based perturbative expansion. In the weak noise limit, the noise-averaged expectation value $\overline{\braket{O(T)}}=\overline{\Tr{[\rho(T)O]}}$ with respect to the time-evolved state $\rho(T)=U(T)\rho(0)U^\dagger(T)$ can be approximated as
\begin{equation}
    \overline{\braket{O(T)}} \approx \Tr{[e^{-i\mathcal{C}_O^{(1)}(T) + \mathcal{C}_O^{(2)}(T)/2}\rho_C(T)O]}.
    \label{eq:exp-val}
\end{equation}
$\rho_C(T)=U_C(T)\rho(0)U^\dagger_C(T)$ represents the time-evolved state with respect to the ideal control dynamics~\cite{Paz-Silva2017QNS, quiroz2021qaoa}, while the error dynamics generated by $H_E(t)$ are described by the first and second cumulants:
\begin{eqnarray}
    \mathcal{C}^{(1)}_O(T) &=&
    \sum_{i< j}^{N} \sum_{\mu,\nu= x,y,z} \chi_{ij}^{\mu\nu}(T) (\sigma^\mu_i \sigma^\nu_{j} - O^{-1}\sigma^\mu_i \sigma^\nu_{j} O)\quad
    \label{eq:first_cumulant}\\
    \mathcal{C}_O^{(2)}(T) &=& 
    \sum_{i,j=1}^{N} \sum_{\nu,\gamma=x,y,z} \Gamma^{\nu\gamma}_{ij}(T) \mathcal{A}^{\nu\gamma}_{ij},
    \label{eq:second_cumulant}
\end{eqnarray}
The cumulants are determined by the operator $\mathcal{A}_{ij}^{\nu\gamma}=\sigma^\nu_i \sigma^\gamma_{j} + O^{-1} \sigma^\nu_{i}\sigma^\gamma_j O  - O^{-1} \sigma^\nu_i O \sigma^\gamma_{j} - \sigma^\nu_i O^{-1} \sigma^\gamma_{j}O$ and the overlap integrals 
\begin{eqnarray}
    \chi_{ij}^{\mu\nu}(T) &\equiv&  J_{ij} \int_0^T R^{z\mu}_i(t)R^{z\nu}_{j}(t) dt, 
    \label{eq:first_overlap} \\
    \Gamma^{\nu\gamma}_{ij}(T) &\equiv& \sum_{\mu,\delta=x,y,z}
     \int_{0}^{\infty}\frac{d\omega}{2\pi}  \mathcal{G}_{ij}^{\mu\nu\delta\gamma}(\omega,T)S^{\mu\delta}_{ij}(\omega) ,
    \label{eq:second_overlap}
\end{eqnarray}
with $S^{\mu\nu}_{ij}(\omega)= \int_0^T C^{\mu\nu}_{ij}(\tau) e^{-i\omega \tau}dt$ designating the noise power spectral density. The filter functions $\mathcal{G}_{ij}^{\mu\nu\delta\gamma}(\omega,T)\equiv{\rm Re}\left[ G^{\mu\nu}_i(\omega,T)G^{\delta\gamma}_j(-\omega,T)\right]$ are defined in terms of the Fourier transforms of the elements of the control matrix: $G^{\mu\nu}_i(\omega,T)\equiv\int^{T}_0 R^{\mu\nu}_i(t)e^{i\omega t}dt$. Note that the first cumulant is solely composed of crosstalk contributions, while the second cumulant only contains system-environment interactions.

Similar expressions can be obtained for the fidelity (i.e., overlap) between the initial state $\rho(0)$ and time-evolved state $\rho(T)$. When $\rho(0)$ is a pure state, the noise-averaged fidelity is given by $\mathcal{F}(T)=\overline{\Tr{[\rho(T)\rho(0)]}}$. In general, $\rho(0)$ can be non-invertible which poses challenges for recasting $\mathcal{F}(T)$ in the form of Eq.~(\ref{eq:exp-val})~\cite{supp}. However, by expanding the initial state as a sum of invertible, Hermitian operators: $\rho(0)=\sum_\ell \Phi_\ell$, the fidelity can be approximated as
\begin{equation}
    \mathcal{F}(T)\approx\sum_\ell\Tr{[e^{-i\mathcal{C}_{\Phi_\ell}^{(1)}(T) + \mathcal{C}_{\Phi_\ell}^{(2)}(T)/2}\rho_C(T)\Phi_\ell]}.
    \label{eq:fidelity}
\end{equation}
The expansion of $\rho(0)$ is particularly applicable for qubit systems, where $\Phi_i$ may be chosen to be proportional to the $N$-qubit Pauli operators. While the cumulant expressions of Eq.~(\ref{eq:fidelity}) differ from those of Eq.~(\ref{eq:exp-val}) in the operator conjugations, their commonality lies in the overlap integrals $\chi^{\mu\nu}_{ij}(T)$ and $\Gamma^{\nu\gamma}_{ij}(T)$. These quantities are exclusively dependent upon the specifications of the noise model and the control.

%
%
\emph{Crosstalk-Robust Control Condition.} -- 
In the cumulant representation, a control protocol defined by $\{\Omega_i(t),\phi_i(t)\}^N_{i=1}$ attains $n$th order suppression of the error dynamics when $\mathcal{C}^{(k)}_O(T)=0$ for $k\leq n$. Suppression of the first cumulant is achieved by cancellation of the pure crosstalk contribution, or more specifically, satisfying 
the \emph{crosstalk suppression condition}:
\begin{equation}
    \chi^{\mu\nu}_{ij}(T)=0\quad \forall i,j,\mu,\nu.
    \label{eq:csc}
\end{equation}
Due to the static nature of $J_{ij}$, $\mathcal{C}^{(2)}_O(T)$ is completely dominated by system-environment interactions. As such, imposing Eq.~(\ref{eq:csc}) in conjunction with minimizing the spectral overlap between the system-environment noise and the filter functions leads to second order suppression of all noise contributions.

The efficacy of the above condition is explored in the next sections via experimental realizations of quantum state preservation and QNS on the IBMQE. Composed of fixed-frequency transmons, IBMQE processors are susceptible to parasitic $ZZ$-crosstalk~\cite{Gambetta2012benchmarking, Ash-Saki2020crosstalk, Vinay2021DD} and temporally-correlated noise~\cite{Morris2019non-Markov,Chen2020non-Markov,Vinay2021DD, White2020non-Markov,zhang2022nonmarkov}, both of which are captured by Eq.~(\ref{eq:H-err}). Thus, they present  suitable testbeds for evaluating and showcasing the potential impact of imposing Eq.~(\ref{eq:csc}).

%
%
\emph{Quantum State Preservation} -- 
We demonstrate the utility of the above condition in the context of quantum state preservation through the design and experimental evaluation of CRDD. DD selectively averages out unwanted interactions between a quantum system and its environment and combats systematic errors through the use of fast and strong pulses~\cite{lidar2013qec-book}.
When properly designed, DD preserves the state of a single qubit system, while suppressing static quantum crosstalk~\cite{Vinay2021DD}. Here, we show how the above condition enables greater generality for simultaneous preservation of an array of qubits initialized in SPSs and MESs.

While the crosstalk suppression condition allows for a diverse family of possible CRDD protocols, for concreteness, we focus on those inspired by \xy4~\cite{viola1999dd}. This sequence utilizes repetitions of $f_\tau X f_\tau Y f_\tau X f_\tau Y$, where $X$ and $Y$ are $\pi$-pulses (of duration $\delta$) about the $x$ and $y$ axis of the single qubit Bloch sphere, respectively. Free evolution periods $f_\tau$, where the system evolves according to its internal dynamics, are of duration $\tau$; thus, yielding a cycle time of $t_{\xy4}=4(\tau+\delta)$. \xy4 affords suppression of system-environment interactions to first order in time-dependent perturbation theory~\cite{viola1999dd}.

\begin{figure}[t]
    \centering
    \includegraphics[width=\columnwidth]{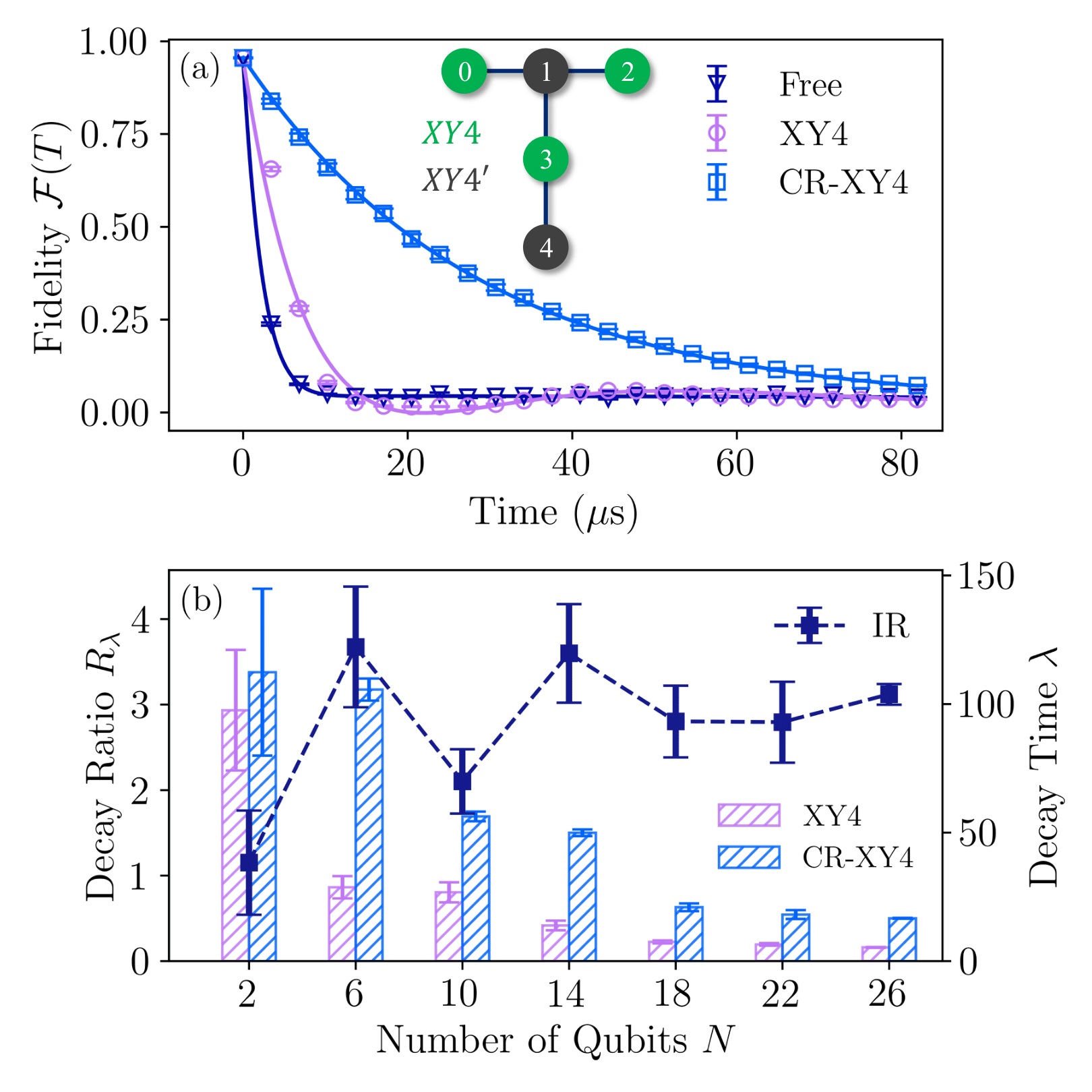}
    \caption{Simultaneous preservation of SPSs using different control protocols. (a) Fidelity vs. time for the IBQME Lima 5-qubit processor using free evolution (dark blue down-triangles), \xy4 (light purple circles), and \crxy4 (light blue squares). Data points and error bars denote mean fidelity and CIs, respectively, obtained from bootstrapping. Inset: \crxy4 protocol on IBMQE Lima; similar patterning used for IBMQE Auckland. (b) Fidelity decay rate comparison between \xy4 and \crxy4 using up to 26 qubits on IBMQE Auckland. Bars represent mean decay time $\lambda$ (with CI error bars), while the line signifies the IR. Results are collected using the same procedure outlined for the top panel. In both cases, \crxy4 exhibits significant improvement over \xy4.}
    \label{fig:sps-dd}
\end{figure}

Crosstalk-robust state preservation is achieved by properly adjusting the pulse locations of multi-qubit variants of \xy4. Given an array of qubits, the suppression condition is enforced by patterning two sequences across the array: \xy4 and \xyp$=Xf_\tau Y f_\tau X f_\tau Y f_\tau$; this protocol is labeled \crxy4. An example of the protocol is shown in the inset of Fig.~\ref{fig:sps-dd}(a). In the ideal, instantaneous pulse limit, simultaneous application of these sequences results in universal single qubit decoupling and proper tailoring of the control matrix to suppress crosstalk. In the DD context, the suppression condition is equivalent to enforcing the typical anti-commutativity required to achieve first-order decoupling~\cite{lidar2013qec-book}. As such, variants of CRDD based on time-symmetric \xy4~\cite{souza2012symdd} and Eulerian DD~\cite{wocjan2006edd} may be considered as well.

\crxy4 is experimentally evaluated against free evolution and \xy4. The three protocols are compared by simultaneous preparing $N$ qubits in the state $\ket{\Psi}=\otimes^{N}_{i=1}\ket{\psi_i}$, where $\ket{\psi_i}$ is a state in the $xy$-plane of the single-qubit Bloch sphere. Then, $M$ repetitions (or the free evolution equivalent) of the protocol are applied for a total time $T=M t_{\xy4}=M t_{\crxy4}$. For all devices discussed below, $\tau=\delta=35$ns. The experiment is completed by applying the inverse state preparation unitary followed by a measurement in the computational basis. Resulting measurements are used to estimate the fidelity $\mathcal{F}(T)$ between $\rho(0)=\ket{\Psi}\bra{\Psi}$ and $\rho(T)$ resulting from free or DD evolution. We focus on states in the $xy$-plane as these are among the most susceptible to $ZZ$-crosstalk.

\crxy4 substantially improves the fidelity of SPSs. In Fig.~\ref{fig:sps-dd}(a), a comparison between the protocols is shown for the IBMQE Lima 5-qubit processor. Estimates of average fidelity and 95\% confidence intervals (CIs) are determined via bootstrapping (with replacement)~\cite{stine1989bootstrap} from 1000 resamples of data collected from 20 random qubit states in the $xy$-plane, 8000 shots, and four replicates of the experiment run over four days. The data is fit to the modified exponential decay~\cite{Pokharel2018DD}: $F(t) = c[1-f(t)] + F_0$, where $f(t) = 1/(1+k)(ke^{-t/\lambda}\cos(t\gamma) + e^{-t/\alpha})$, and $c = (F_{T_{\max}} - F_0)/(f(T_{\max})-1)$.
The model includes short and long decay times $\lambda$ and $\alpha$, respectively, oscillation frequency $\gamma$, and dimensionless weight parameter $k$. The calculated fidelity at $M=0$ and $M=M_{\max}$ are given by $F_0$ and $F_{T_{\max}}$, respectively. A comparison of the short decay times via the ratio $R_\lambda=\lambda_{\crxy4}/\lambda_{\xy4}$ reveals a factor of 4 improvement for \crxy4 over \xy4. While the enhancement varies as the number of qubits increases due to hardware variability, it remains near a $3\times$ improvement up to 26 qubits, as shown experimentally on IBMQE Auckland in Fig.~\ref{fig:sps-dd}(b).

\begin{figure}[t]
    \centering
    \includegraphics[width=\columnwidth]{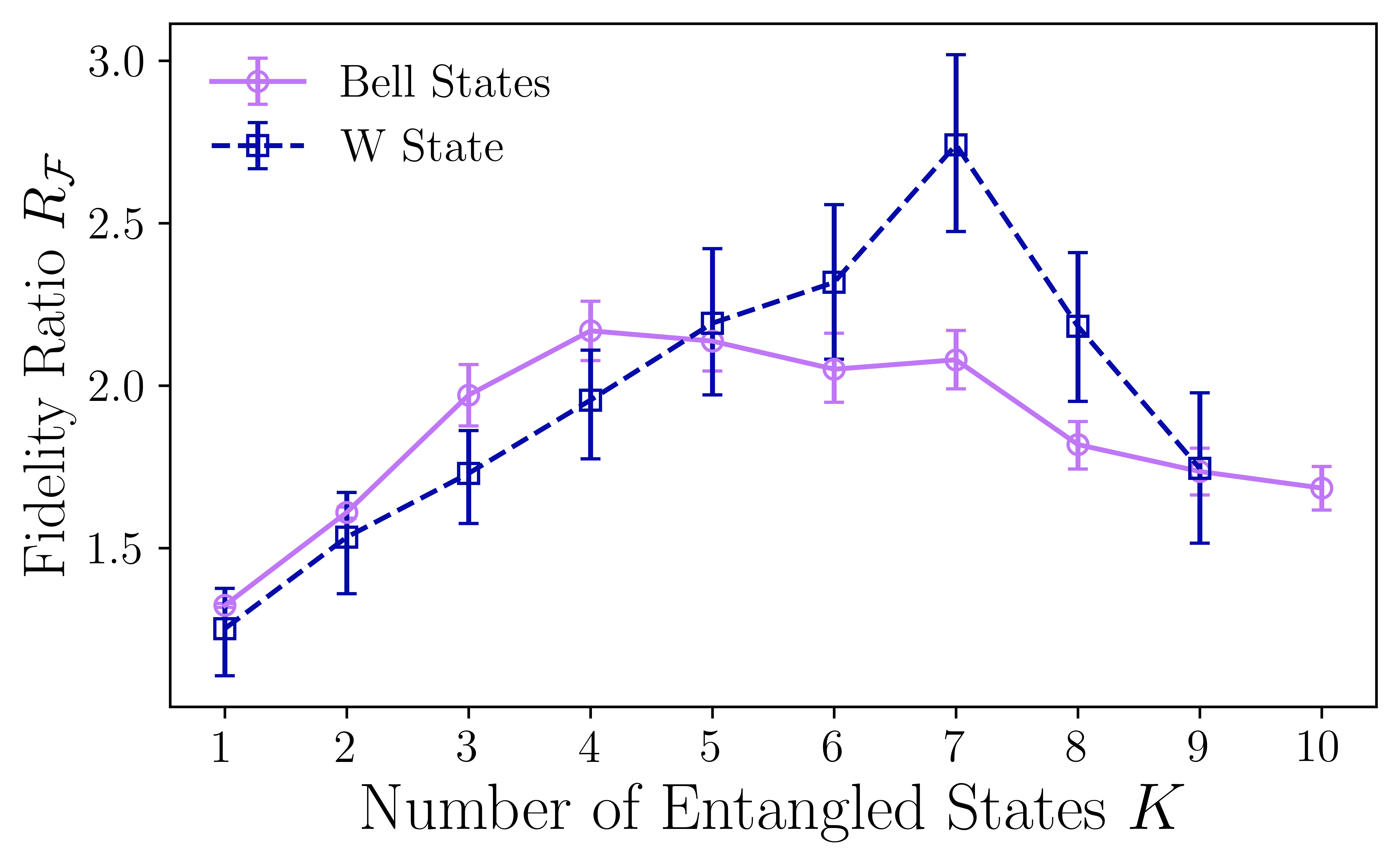}
    \caption{Ratio of time-averaged fidelities of \crxy4 and \xy4 for the simultaneous protection of $K$ MESs on the 27-qubit IBMQE Auckland processor. Plot contains results for Bell states (light purple circles) and the $W$ state (dark blue squares). Bell state results are collected using 8000 shots, the four Bell states, and five replicates of the experiment collected over five days. Similar data is collected for the $W$ state. Averages and CIs are determined by bootstrapping, where 1000 resamples of the data are used. Results indicate that \crxy4 improves state preservation over \xy4 for both cases.}
    \label{fig:mes-scaling}
\end{figure}

\crxy4 considerably enhances the time-average fidelity of MESs. In Fig.~\ref{fig:mes-scaling}, $N$ qubits are simultaneously prepared in $K^\prime$ entangled states and then subject to $M$ repetitions of \xy4 or \crxy4. Then, prior to measurement in the computational basis, the inverse state preparation unitary is applied. Upon applying up to $M_{\max}=50$ repetitions of DD, the time-averaged fidelity $\mathcal{F}_{\rm avg}= T^{-1}_{\max}\int_0^{T_{\max}} \frac{\mathcal{F}(t)}{\mathcal{F}(0)}dt$ is calculated via numerical integration for each DD protocol~\cite{ezzell2022dynamical}. Note that this measure captures long-time behavior, with the normalization accounting for state preparation errors. In Fig.~\ref{fig:mes-scaling}, $\mathcal{F}_{\rm avg}$ is conditioned on the simultaneous preservation of $K\leq K^\prime$ states on physically adjacent qubits. We consider $N=20$ qubits simultaneously prepared in $K^\prime=10$ Bell states of the form $\ket{\Phi_{\pm}}=1/\sqrt{2}(\ket{00}\pm\ket{11})$ or $\ket{\Psi_{\pm}}=1/\sqrt{2}(\ket{01}\pm\ket{10})$. 
A similar procedure is used for the three-qubit $W$ state $\ket{W}=1/\sqrt{3}(\ket{001}+\ket{010}+\ket{100})$, where $K^\prime=9$ entangled states are prepared on $N=27$ qubits. In Fig.~\ref{fig:mes-scaling}, the fidelity ratio $R_\mathcal{F}=\mathcal{F}^{\crxy4}_{\rm avg}/\mathcal{F}^{\xy4}_{\rm avg}$ is shown for both state preparations. $\mathcal{F}_{\rm avg}$ collected from all initial states, 8000 shots, and five replicates of the experiments are used to estimate $R_\mathcal{F}$ and CIs via bootstrapping~\cite{supp}. Individual Bell states are invariant under $ZZ$-crosstalk, however, multiple Bell states prepared physically adjacent on the quantum device experience $ZZ$-crosstalk across the common edge. The impact of suppressing edge effects is observed via the $2\times$ improvement over $\xy4$ obtained by \crxy4. The $W$ state does not possess inherent robustness and therefore experiences a greater benefit from CRDD; a near $3\times$ improvement. Despite the increasing contributions from state preparation and measurement and gate error observed at large $N$, \crxy4 continues to achieve a slower decay in fidelity than \xy4 and hence, a higher time-averaged fidelity for both MES preparations considered.

\begin{figure}[t]
    \centering
    \includegraphics[width=\columnwidth]{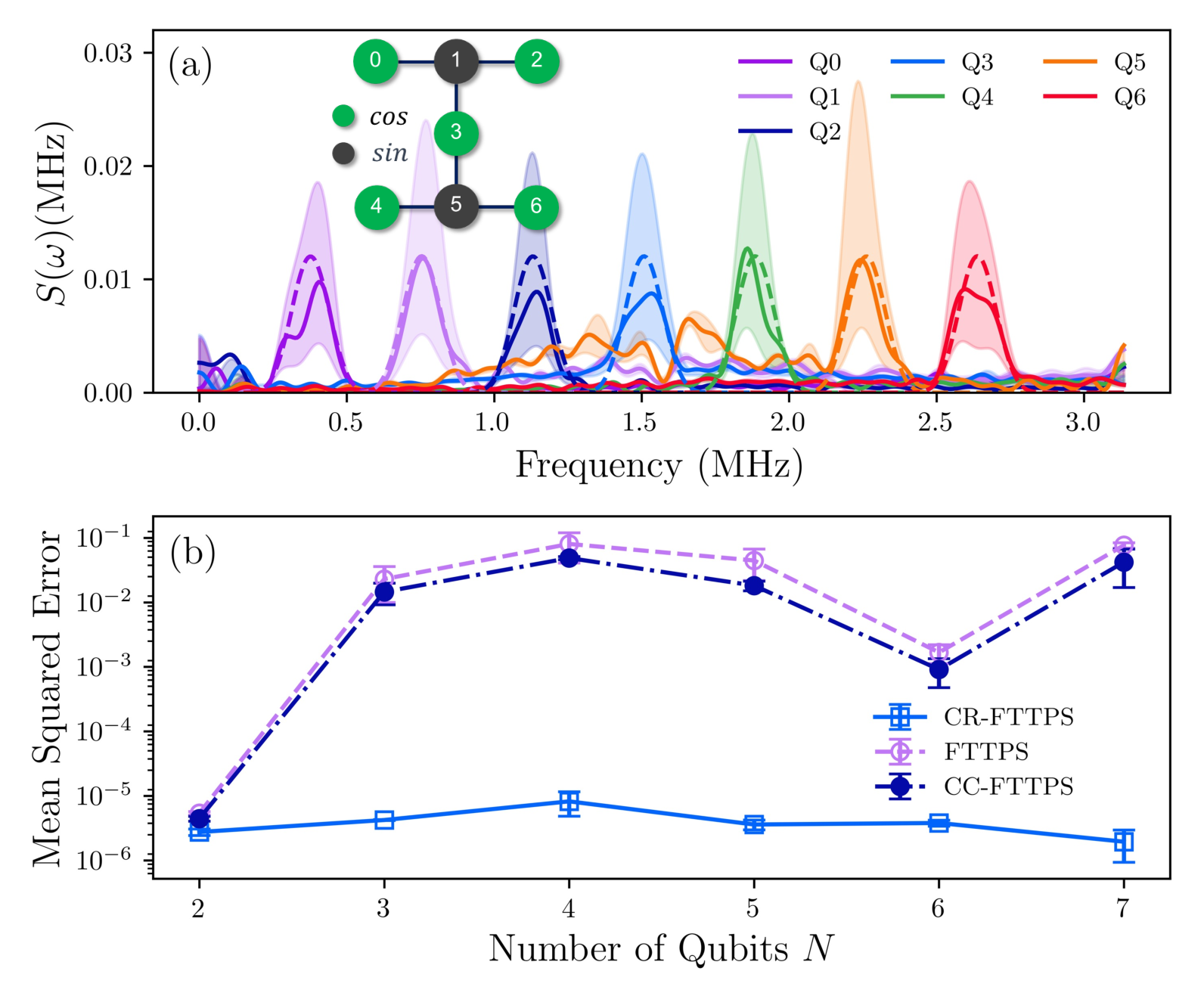}
    \caption{Crosstalk-robust QNS on the 7-qubit IBMQE Nairobi processor, where narrowband dephasing noise with distinct center frequencies is injected on each qubit. (a) Seven-qubit simultaneous local dephasing QNS using CR-FTTPS. Average spectrum estimates (solid lines) and CIs (shaded regions) indicate good agreement with injected noise (dashed lines). Inset: CR-FTTPS protocol on Nairobi. (b) Reconstruction error as a function of the number of qubits for FTTPS variants. Data points denote averages and error bars are CIs.  CR-FTTPS outperforms other variants considered by up to 4 orders of magnitude. Averages and CIs are determined from bootstrapping.}
    \label{fig:qns}
\end{figure}

\emph{Quantum Noise Spectroscopy} -- The crosstalk suppression condition can inform the design of QNS protocols. In QNS, a controlled quantum system is used as a dynamical probe to characterize the spectral properties of environmental noise. Under the zero-mean noise assumption, spectrum estimates are determined by tailoring the frequency response of the system via control, estimating expectation value decay rates, and using $\mathcal{C}^{(2)}_O(T)$ to define a linear inversion problem at discrete frequencies~\cite{alvarez2011qns,bylander2011qns}. Crosstalk introduces spurious features in spectral estimates that are further exacerbated by imperfections in the control~\cite{supp}. We construct pulse-based sequences that provide robustness to both crosstalk and errors due to finite pulse duration while enabling estimates of local dephasing noise spectra on a collection of qubits.

In particular, we design a crosstalk-robust variant of the Fixed Total Time Pulse Sequences (FTTPS)~\cite{Schultz2021schwarma,Murphy2022schwarma}. Bookended by two $X_{\pi/2}$ pulses, FTTPS consist of $\ell/2$ different sequences, each containing $\ell$ gates. In the instantaneous pulse limit, ``cosine" FTTPS yield the discrete-time control matrix $R^{zz}_i(t)=\sgn\{\cos(\pi (\kappa-1)m/\ell)\}$ for $\kappa,m=1,\ldots, \ell/2$. Sign changes in the cosine function denote locations of $X$ gates and $I$ gates otherwise. Simultaneous application of ``cosine'' FTTPS on multiple qubits preserves the first cumulant and therefore crosstalk. However, by introducing an additional ``sine" FTTPS described by $R^{zz}_i(t)=\sgn\{\sin(\pi (\kappa-1)m/\ell)\}$ alternating the qubit array with the cosine and sine variants, one can effectively implement crosstalk-robust FTTPS (CR-FTTPS).

Using the IBMQE Nairobi 7-qubit processor, we inject narrow band dephasing noise at $N$ distinct frequencies on $N$ qubits via the Schrodinger Wave Moving Average Model (SchWARMA) approach~\cite{Murphy2022schwarma}. The spectrum reconstruction accuracy of CR-FTTPS and FTTPS for $\ell=128$ is assessed using 10 SchWARMA trajectories and five data sets collected over five days. Bootstrapped spectrum estimates and CIs are determined from 1000 resamples of the reconstructed spectra. Figure~\ref{fig:qns}(a) shows spectrum estimates using CR-FTTPS on 7 qubits, where average spectrum estimates (solid lines) and CIs (shaded regions) agree well with the injected spectra (dotted lines). Analyzing the mean-squared error between estimated and injected spectra as a function of the number of qubits [Fig.~\ref{fig:qns}(b)], we find that CR-FTTPS achieves a $10^4$ improvement over FTTPS in reconstruction accuracy. This enhancement is due to suppression of $ZZ$, as well as pulse-error-induced $ZY$ and $YZ$-crosstalk. We substantiate this claim empirically by considering crosstalk-corrected FTTPS (CC-FTTPS), where $ZZ$-crosstalk is included in the spectrum reconstruction by using IBM-measured crosstalk coupling strengths; only slight improvements are observed. CR-FTTPS robustness is further justified analytically in the supplement.

\emph{Conclusions} -- Through the lens of quantum control, we develop a condition for first-order quantum crosstalk suppression for general single qubit control of multi-qubit systems. The utility of the condition is demonstrated in the domains of quantum state preservation and noise characterization, where we design crosstalk robust DD and QNS, respectively. Through experimental investigations on the IBMQE, we showcase the significance of the condition in each domain on various processors, simultaneously mitigating crosstalk on up to 27 qubits. Together, these experiments highlight the crucial impact our condition can have on improving simultaneous characterization and control on current quantum devices.

\begin{acknowledgments}
\emph{Acknowledgements} -- This work was supported by the U.S. Department of Energy, Office of Science, Office of Advanced Scientific Computing Research, Accelerated Research in Quantum Computing under Award Number DE-SC0020316. KS and GQ acknowledge support from ARO MURI grant W911NF-18-1-0218. This research used resources of the Oak Ridge Leadership Computing Facility, which is a DOE Office of Science User Facility supported under Contract DE-AC05-00OR22725.
\end{acknowledgments}

\appendix
\section{Derivation of the Effective Error Dynamics}
\subsection{Rotated-Frame Dynamics}
Time-dependent perturbation theory (TDPT) is a useful tool to evaluate the dynamics of a quantum system in the presence of weak noise and strong control. TDPT usually involves moving the noise Hamiltonian from the lab frame to the rotating frame with respect to the control, also known as the ``toggling frame". The error dynamics are then evaluated through perturbative expansion methods (e.g. cumulant expansion). Here, we study the error dynamics in a ``reverse" toggling frame representation as it is convenient for isolating the error dynamics from the ideal state evolution. The error Hamiltonian in this toggling frame is specified by
\begin{equation}
    \tilde{H}_E(t) = U_C(T,t)H_E(t)U_C^{\dagger}(T,t)\label{rev-f}
\end{equation}
where $U_C(T,0)= \mathcal{T}_+e^{-i\int_0^T H_C(s)ds}$ denotes the ideal control operator. The total dynamics of the quantum system $U(T) = \mathcal{T}_+ e^{-i\int_0^TH(t)dt}$ can be factorized into the product of the rotated error operator and ideal control operator according to $U(T) = \tilde{U}_E(T)U_C(T)$. The rotated error operator $\tilde{U}_E(T)$ is generated by the rotated-frame Hamiltonian in Eq.~\ref{rev-f}, i.e., $\tilde{U}_E(T)=\mathcal{T}e^{-i\int_0^T \tilde{H}_E(s)ds}$. Note that the alternative toggling frame representation can be equivalently transformed to the canonical toggling frame representation via
\begin{eqnarray}
    U(T) &=& U_C(T)\tilde{U}'_E(T) \label{c-rot-frame} \nonumber\\ 
    &=& U_C(T)\tilde{U}'_E(T)U_C^{\dagger}(T)U_C(T) \label{rot-transform} \nonumber\\ 
    &=& 
    \tilde{U}_E(T)U_C(T).
\end{eqnarray}
Eq.~\ref{c-rot-frame} represents the canonical interaction picture representation with $\tilde{U}'_E(T) = \mathcal{T}_+\exp[-i\int_0^T \tilde{H}'_E(t)dt]$, and the generating Hamiltonian $\tilde{H}'_E(t)$ can be expressed $\tilde{H}'_E(t) = U_C^{\dagger}(T,0)H_E(t)U_C(T,0)$. The two toggling frame representations are equivalent, differing by the ordering of the two evolution operators and the time-evolution-generating error Hamiltonians. The utility of the reverse interaction picture becomes evident in Sec.~\ref{subsec:obs-dyn}.

\subsection{Time-dependent Dynamics of Observables}
\label{subsec:obs-dyn}
The rotated frame discussed in the previous section can be a useful tool for examining the effect of noise on the expectation value of observable $O$. We focus on the noise-averaged time-dependent expectation value described by
\begin{equation}
    \overline{\langle O(T)\rangle} = \overline{\langle \Tr[\rho(T)O(T)]\rangle},
\end{equation}
where $\rho(T)$ is the density matrix representing the system's state at time $T$. By partitioning the overall operator $U(T)$ into $\tilde{U}_E(T)U_C(T)$ as shown in the previous section, the density matrix can be expressed as
\begin{eqnarray}
    \rho(T) &=&
    U(T)\rho(0)U^{\dagger}(T) \nonumber\\
    &=& \tilde{U}_E(T)U_C(T)\rho(0)U_C^{\dagger}(T)\tilde{U}_E^{\dagger}(T) \nonumber\\
    &=& \tilde{U}_E(T)\rho_C(T)\tilde{U}_E^{\dagger}(T),
\end{eqnarray}
with $\rho_C(T) = U_C(T)\rho(0)U_C^{\dagger}(T)$. The expectation value can be rewritten as
\begin{eqnarray}
    \overline{\langle O(T)\rangle} &=& \overline{\langle \Tr[\tilde{U}_E(T) \rho_C(T) \tilde{U}_E^{\dagger}(T)O]\rangle} \nonumber\\
    &=& \Tr[\Theta(T)\rho_C(T)O] 
\end{eqnarray}
This allows the total dynamics to be partitioned into the error operator $\Theta(T)=\overline{O^{-1}\tilde{U}_E^{\dagger}(T)O\tilde{U}_E(T)}$ and unperturbed time-evolved states $\rho_C(T)$. The error dynamics are captured within one single function $\Theta(T)$, which can be further evaluated using a cumulant expansion.

\subsection{Cumulant Expansion}
The cumulant expansion is a moment expansion approach that characterizes the probability distribution of a certain quantity. Originally from probability theory, cumulant expansions can be useful for evaluating the error dynamics when the error Hamiltonian contains stochastic variables. 
\par
Here, the utility of the cumulant expansion emerges by first writting the error propagator $\Theta(T)$ as
\begin{equation}
    \Theta(T) = \overline{\mathcal{T}_+e^{-i\int_{-T}^T\tilde{H}_{E,O}(t)dt}}
\end{equation}
with observable-dependent effective Hamiltonian $\tilde{H}_{E,O}(t)$
\begin{equation}
  \tilde{H}_{E,O}(t) =
    \begin{cases}
      \tilde{H}_E(T+t) & t \in [-T,0]\\
      -O^{-1}\tilde{H}_E(T-t)O & t \in [0,T]
    \end{cases}       
\end{equation}
We can expand error operator according to $\Theta(T) = e^{\mathcal{C}_O(T)}$ using moment-generating function with $\mathcal{C}_O(T) = \sum\limits_{n=1}^{\infty}(-i)^{n}\frac{\mathcal{C}_O^{(n)}(T)}{n!}$. While the cumulant expansion yields an infinite series, we adopt a second order truncation as a sufficient approximation to the error dynamics. More concretely, the second order truncation is valid if
\begin{equation}
    || \mathcal{C}_O(T)||_{\infty} \simeq
    ||-i\mathcal{C}^{(1)}_O(T) + \frac{\mathcal{C}^{(2)}_O(T)}{2!}||_{\infty},
\end{equation}
where $\|\cdot\|_\infty$ denotes the operator norm or largest singular value. The truncated cumulant expansion gives rise to Eq.~(4) in the main text. By the change of effective Hamiltonian to observable-independent Hamiltonian and the change of integral boundaries, we obtain an explicit expression of the first and second order cumulants
\begin{widetext}
\begin{eqnarray}
    \mathcal{C}^{(1)}_O(T) &=& \int_{-T}^{T}dt_1 \langle \tilde{H}_{E,O}(t_1)\rangle = \int_0^T dt_1 [\langle \tilde{H}_E(t_1)\rangle - \langle O^{-1}\tilde{H}_E(t_1)O\rangle]\\
    \mathcal{C}_O^{(2)}(T) &=&\int_{-T}^{T}dt_1\int_{-T}^{T}dt_2 \braket{\tilde{H}_{E,O}(t_1)\tilde{H}_{E,O}(t_2)} - \int_{-T}^{T}dt_1\braket{\tilde{H}_{E,O}(t_1)}
     \int_{-T}^{T}dt_2\braket{\tilde{H}_{E,O}(t_2)}\nonumber\\
     &=& \int_{0}^{T}dt_1\int_{0}^{T}dt_2 \left(\braket{\tilde{H}_{E}(t_1)\tilde{H}_{E}(t_2)} + \braket{\overline{\tilde{H}}_{E}(t_{1})\overline{\tilde{H}}_{E}(t_{2})O} - \braket{\overline{\tilde{H}}_{E}(t_1)\tilde{H}_{E}(t_2)} - \braket{\tilde{H}_{E}(t_1)\overline{\tilde{H}}_{E}(t_2)}\right) \nonumber\\
     && - \left(\int_0^T dt\, [\braket{\tilde{H}_E(t)} - \braket{O^{-1}\tilde{H}_E(t_1)O}]\right)^2,
\end{eqnarray}
\end{widetext}
where $\overline{\tilde{H}}_E(t) = O^{-1}\tilde{H}_E(t)O$.
The results of $\mathcal{C}_O^{(1)}(t)$ and $\mathcal{C}_O^{(2)}(t)$ for the specific model considered in this study are presented in Eqs.(7)-(8) in the main text. The first cumulant is solely composed of crosstalk contributions, while the second order cumulant only contains system-environment interactions by explicitly evaluating each sub-terms of the first two orders of cumulant. This same conclusion can be achieved from a purely statistical point of view as well. By the definition of a cumulant series, the first-order cumulant represents the process mean, and the second-order cumulant represents the process variance. In our noise model, semi-classical spatio-temporally correlated noise processes is time-dependent but with a zero mean, $\overline{\beta_i^{\mu}(t)} = 0$. Static quantum crosstalk has a non-zero mean but a zero variance. As such, the statistical properties of the two types of noise automatically give rise to results that the semi-classical system-environment noise has no contribution to the first cumulant, and the static quantum crosstalk has no contribution to the second cumulant. Note that the second order cumulant, following the canonical FFF, can be written in the frequency domain in terms of the so-called overlap integral \cite{Green2013control}.

\subsection{Noise-Averaged Fidelity}
In the main text, the noise-averaged expectation value is accompanied by the noise-averaged fidelity. When the system is initially in a pure state, the latter can be formally expressed as $\mathcal{F}(T)=\overline{\Tr{[\rho(T)\rho(0)]}}$. This expression can be written in terms of the time-evolved state with respect to the ideal control dynamics following the same procedure as in Sec.~\ref{subsec:obs-dyn}. Namely, let $\rho(0)=\sum_\ell \Phi_\ell$, where $\Phi_\ell$ are invertible, Hermitian operators. The fidelity is then given by
\begin{eqnarray}
    \mathcal{F}(T) &=& \sum_\ell\overline{\Tr{[\rho(T)\Phi_\ell]}} \nonumber\\
    &\stackrel{(1)}{=}& \sum_\ell\overline{\Tr{[\tilde{U}_E(T)\rho_C(T)\tilde{U}^\dagger_E(T)\Phi_\ell]}}\nonumber\\
    &\stackrel{(2)}{=}& \sum_\ell\Tr{[\overline{\Phi^{-1}_\ell\tilde{U}^\dagger_E(T)\Phi_\ell\tilde{U}_E(T)}\rho_C(T)\Phi_\ell]}\nonumber\\
    &=&\sum_\ell\Tr{[\Theta_\ell(T)\rho_C(T)\Phi_\ell]},
\end{eqnarray}
where (1) follows from the factorization of the total time evolution operator and (2) follows from the cyclicity of the trace and the introduction of $\Phi^{-1}\Phi$. Within the final expression is the error operator $\Theta(T)_\ell=\overline{\Phi^{-1}_\ell\tilde{U}^\dagger_E(T)\Phi_\ell\tilde{U}_E(T)}$ that can be expressed as a cumulant expansion. Enforcing the weak noise assumption leads to the second-order truncated expression given in the main text.

\section{General Quantum Crosstalk Suppression Criteria}
In the main text of the paper, we focus on the suppression of static $ZZ$-interaction as the primary form of quantum crosstalk. In near-term devices, however, other types of quantum crosstalk also prevail in various qubit architectures \cite{ospelkaus2008trapped, urban2009rydberg, Auger2017Rydberg, Levine2018Rydberg, Parrado-Rodriguez2021trappedion,Fang2022trapped-ion}. For example, $XX$-interaction proves to dominate quantum crosstalk in trapped-ion qubits~\cite{Zhang2022inverse}. In this section, we present the generalization of the \emph{crosstalk suppression conditions} for multi-axis quantum crosstalk noise and single-qubit control. The qubit noise is generated by the error Hamiltonian \begin{equation}
    H_E(t) = \sum\limits_{i=1}^{N}\vec{\sigma_{i}}\cdot\vec{\beta_i}(t) + \sum^N_{i< j, \alpha, \beta} J^{\alpha,\beta}_{ij}\,\sigma_i^{\alpha}\sigma_j^{\beta},
    \label{eq:H-err-gen}
\end{equation}
with $\alpha,\beta \in \{x,y,z\}$ denotes all possible two-qubit coupling. With the same expression for control Hamiltonian presented in the main text, the rotated-frame error Hamiltonian is specified by
\begin{eqnarray}
    \tilde{H}_E(t) 
    &=& \sum_{i=1}^N \vec{\Lambda}_i(t) \cdot \vec{\beta}_i(t) \nonumber\\ 
    &&+ \sum_{i,j=1}^N\sum_{\alpha,\beta} J_{ij}^{\alpha,\beta} [\vec{\Lambda}^T_i(t)]_\alpha[\vec{\Lambda}_j(t)]_\beta,\label{eq:H-rot}
\end{eqnarray}
where $\vec{\Lambda}_i(t)\equiv\boldsymbol{R}_i(t) \vec{\sigma}^T_i$ and $\boldsymbol{R}_i(t)$ is the ``control matrix'' with elements ${R}^{\mu\nu}_i(t)=\Tr \left[ U_{C}(T,t)\sigma_{i}^\mu U^{\dagger}_{C}(T,t)\sigma_{i}^\nu \right]/2$. $A^T$ and $[\vec{a}]_\alpha$ denote the transpose of $A$ and the $\alpha$-component of $\vec{a}$, respectively.
The corresponding first order cumulant that encapsulates a multi-axis quantum crosstalk noise is given by 
\begin{equation}
   \mathcal{C}^{(1)}_O(T) =
    \sum_{i< j}^{N} \sum_{\mu,\nu= x,y,z}\sum_{\alpha,\beta} \chi_{ij}^{\mu\nu,\alpha\beta}(T) (\sigma^\mu_i \sigma^\nu_{j} - O^{-1}\sigma^\mu_i \sigma^\nu_{j} O),\quad
    \label{eq:first_cumulant} 
\end{equation}
while the second order cumulant remains unchanged since all crosstalk noise are assumed static. And the first-order overlap integral extends to a $9\times9$ matrix
\begin{equation}
    \chi_{ij}^{\mu\nu,\alpha\beta}(T) \equiv  J^{\alpha\beta}_{ij} \int_0^T R^{\alpha\mu}_i(t)R^{\beta\nu}_{j}(t) dt,
    \label{eq:first_overlap}
\end{equation}
Similarly, imposing condition 
\begin{equation}
    \label{eq:first_order_suppression_criteria}
    \chi_{ij}^{\mu\nu,\alpha\beta}(T) = 0 \quad \forall i,j,\mu,\nu,\alpha,\beta
\end{equation}
enables first-order suppression of multi-axis quantum crosstalk. The generalization extends the suppression of $ZZ$-interaction to the suppression of nine different quantum interactions specified by $\sigma_i^\alpha \sigma_j^\beta$; each type of crosstalk requires a specific quantum control design to mitigate the nine elements in its corresponding control matrix.

\section{Quantum State Preservation} \label{appendix:QSP}

\subsection{Hardware Specification}
\begin{figure}[h]
    \centering
    \includegraphics[width=\columnwidth]{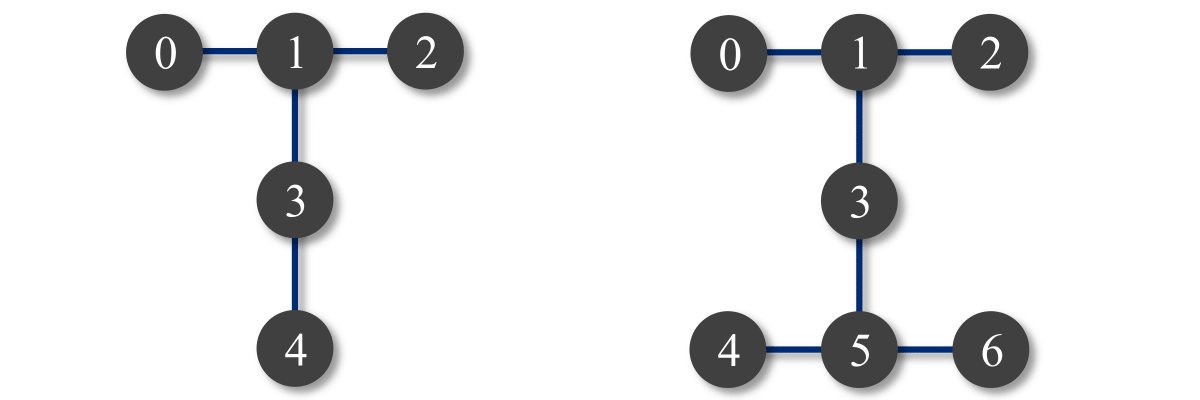}\vspace{0.5cm} \\
     \includegraphics[width=8.5cm]{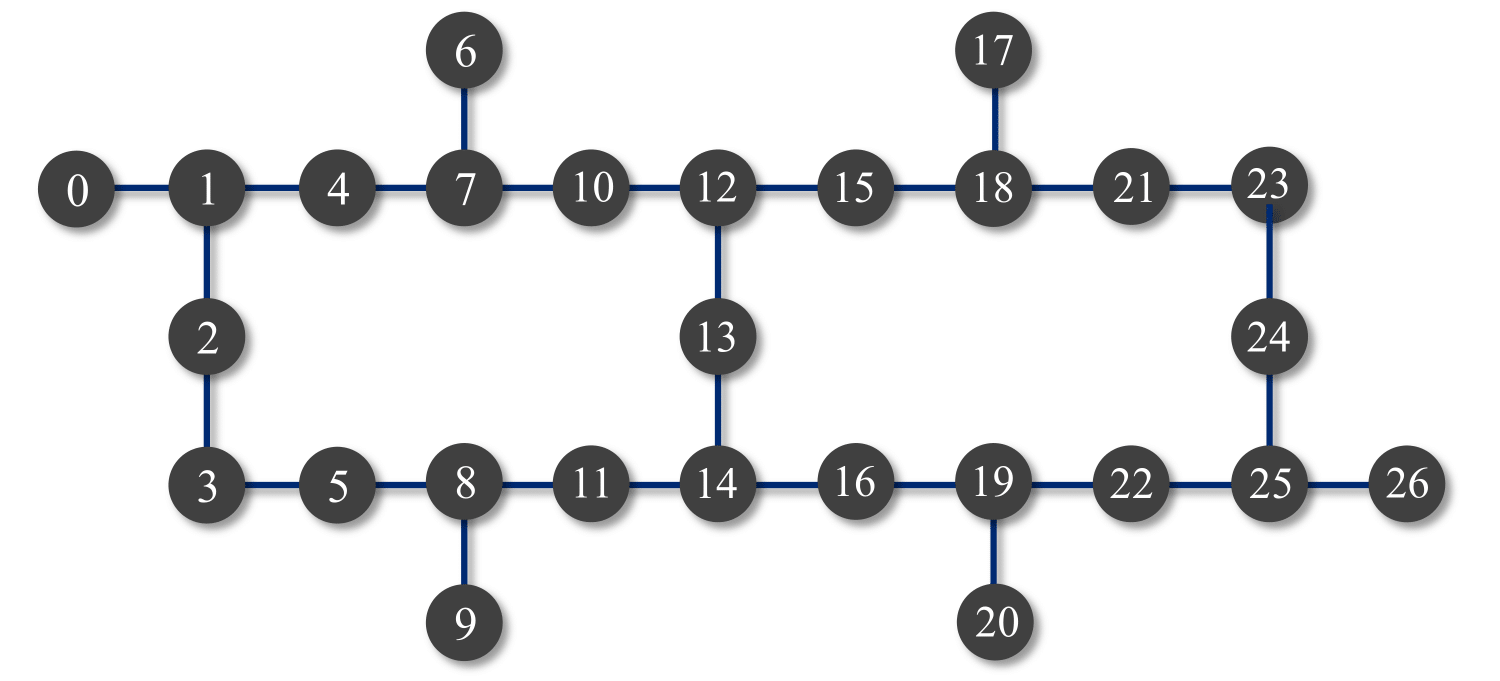}
     \caption{Topologies for devices used to perform quantum state preservation experiments. This includes: the IBM Lima device (top-left), Nairobi (top-right), and the Auckland processor (bottom).}
    \label{fig:lima_nairobi}
\end{figure}

All experiments were performed on IBM Quantum Experience (IBMQE). IBMQE is a cloud-based quantum computing resource that offers access to superconducting transmon quantum processors. All circuits are written in Python using the Qiskit package. The three devices that are used to conduct experimental studies were: IBMQ Lima (5 qubits), IBM Nairobi (7 qubits), and IBM Auckland (27 qubits). The device topology for the processors are shown in Fig.~\ref{fig:lima_nairobi}
The processor, quantum volume (QV), circuit layer operations per seconds (CLOPS), and single-qubit gate duration for each device are detailed in Table~\ref{tab:hardware_specs}.
\begin{table}[h]
    \centering

    \begin{tabular}{cccc}
    \hline\hline
    Device & IBMQ Lima & IBM Nairobi & IBM Auckland \\
    \hline
         Processor & Falcon r4T & Falcon r5.11 H & Falcon r5.11 \\
         QV & 8 & 32 & 64 \\
         CLOPS & 2.7K & 2.6K & 2.4K \\
         1QG Duration & 35.6 ns  & 35.6 ns & 35.6 ns \\
    \hline\hline
    \end{tabular}
    \caption{Device specification for IBMQ Lima, IBM Nairobi, and IBM Auckland. QV, CLOPS, and 1QG denote quantum volume, circuit layer operations per seconds, and 1-qubit gate, respectively.}
    \label{tab:hardware_specs}
\end{table}

\subsection{Data Fitting}
\begin{table}[b]
    \centering
    \begin{tabular}{ccccc}
    \hline\hline
    Evolution & $F_0 \times 10^{-2}$ & $F_{T_{\textrm{max}}} \times 10^{-2}$ & $\lambda$ & $\gamma$ \\ \hline
         Free & $95.5\pm0.1$ &  $4.1\pm0.1$ & $2.21\pm0.05$ & 0 \\
         XY4 & $95.6\pm0.1$ &  $3.5\pm0.1$ & 
         $7.76\pm0.86$ & $0.12\pm0.01$ \\
         CR-XY4 & $95.5\pm0.1$  & $7.2\pm0.2$  &    $28.19\pm0.37$  & 0\\
    \hline\hline
    \end{tabular}
    \caption{Fit parameters when Eq. (\ref{full_mod}) is used to fit Fig. 1(a) in the main text. The decay times $\lambda$ significantly increase when implementing CR-XY4. The modulation frequency $\gamma$ vanishes for free evolution and CR-XY4. }
    \label{tab:fit_param}
\end{table}

\begin{table*}[t]
    \begin{tabular}{cccccccc}
    \hline\hline
    Evolution & $N=2$ & $N=6$ & $N=10$ & $N=14$ & $N=18$ & $N=22$ & $N=26$ \\
    \hline
    XY4 & $97.60\pm11.98$ & $28.74\pm2.22$ & $26.77\pm1.99$ & $13.87\pm0.95$ & $7.48\pm0.30$ & $6.49\pm0.25$ & $5.34\pm0.08$\\
    CR-XY4 & $112.40\pm16.55$ & $105.62\pm2.20$ & $56.28\pm0.95$ & $49.92\pm0.66$ & $20.97\pm0.76$ & $18.13\pm0.87$ & $16.66\pm0.08$\\
    $R_{\lambda}$ & $1.15\pm0.31$ & $3.67\pm0.36$ & $2.10\pm0.19$ & $3.60\pm0.29$ & $2.80\pm0.21$ & $2.79\pm0.24$ & $3.12\pm0.06$ \\
    \hline\hline
    \end{tabular}
    \caption{Decay times $\lambda$ when Eq. \ref{full_mod} is used to fit the fidelity decay curves for $N$ qubits. Decay ratio $R_{\lambda}$ denotes the ratio of $\lambda$ of CR-XY4 to the $\lambda$ of XY4.}
    \label{tab:scaling_decay}
\end{table*}

The fidelity decay curves are fit with a modified exponential decay to quantify the decay rates for qubits subject to different DD protocols. The model is adopted from \cite{Pokharel2018DD}, with an additional coefficient $k$ to adjust the ratio between long- and short-term decay. The coefficient $k$ also proved to be useful in fitting decay curves for qubits subject to quantum crosstalk from more than one neighboring qubit. Here, we find that such qubit dynamics require a model more complex than an exponential decay with oscillation alone. The model consists of four free parameters $\lambda$ (dimensionless long decay times), $\gamma$ (dimensionless modulation frequency), $\alpha$ (dimensionless short decay times), and $k$ (dimensionless adjustment coefficient): 
\begin{eqnarray}
    &&F(t) = cf(t) + c_0,
\end{eqnarray}
where $f(t)$, $c$, and $c_0$ are respectively:
\begin{eqnarray}
    f(t) \,=
    &&\frac{1}{1+k}(e^{-t/\lambda}\cos(\gamma t)+ke^{-t/\alpha}) \\  \label{full_mod}
    c =&& \frac{F_{T_{\textrm{max}}}-F_0}{f(T_{\textrm{max}})-1},\,\, c_0 = F_0 - c
\end{eqnarray}
Here, $F_0$ is the initial fidelity, $F_{T_{\textrm{max}}}$ is the fidelity at time $T_{\max} = 81.92\mu s$, the maximum evolution time when the number of repetitions $N=288$. The corresponding error bars are obtained through bootstrapping detailed in Sec.~\ref{bootstrapping}.

\subsection{Collective Qubit Decay Fit Parameters}
In Fig.~1 in the main text, a collection of qubits were prepared in SPSs, subject to various control protocols, returned to the ground state, and then measured. Here, the focus was the combined qubit state and therefore, the probability that all qubits would return to their ground state at the end of the experiment. In Fig.~1(a), the fidelity decay curve for the set of five qubits returning to their ground states can be sufficiently modeled by an exponential decay with an oscillatory term. As such, a simplified model is used, where the adjustment coefficient $k$ is set to zero; thus, reducing the fit to $f(t) = e^{-t/\lambda}\cos(\gamma t)$. The fit parameters, initial and final fidelities, as well as the error bars are reported in Table \ref{tab:fit_param}. Fit parameters convey that XY4 implemented simultaneously on five qubits improves the fidelity decay time by a factor of three compared to the decay time of free evolution. Furthermore, applying CR-XY4 on five qubits further improves decay times by a factor of four compared with standard XY4. Note that the CR sequence suppresses the oscillation in fidelity caused by quantum crosstalk. In Fig.~1(b), we perform similar experiments and investigate the ratio of decay times between XY4 and CR-XY4 as a function of the number of qubits $N$, increasing from $2$ to $26$ (increment by $4$). The decay ratio denotes the ratio between the decay times of CR-XY4 and XY4 evolution, which quantifies the improvement of CR-XY4 for multi-qubit protection. Experiments are performed on the IBM Auckland device. The decay times and decay ratio are reported in Table \ref{tab:scaling_decay}. 

\begin{table*}[t]
    \centering
    \begin{tabular}{cccccccc}
    \hline\hline
    Qubit & Evolution & $F_0 \times 10^{-2}$ & $F_{T_{\textrm{max}}} \times 10^{-2}$ & $\lambda$ & $\gamma$ & $\alpha$ & $k$\\ 
    \hline
         Q0 & Free & $99.2\pm 0.0$ & $50.8\pm 0.7$ & $9.20\pm 2.51$ & $0.12\pm 0.02$ & $\infty$ & $0$ \\
         Q0 & XY4  & $99.2\pm 0.0$ & $48.4\pm 0.6$ & $36.04\pm 1.70$ & $0.11\pm 0.00$ & $\infty$ & $0$\\
         Q0 & CR-XY4 & $99.2\pm 0.0$ & $59.5\pm 0.6$ & $171.27\pm 10.85$ & $0$ & $\infty$ & $0$\\
         Q0 & XY4 + Idle & $99.2\pm 0.0$ & $61.0\pm 0.5$ & $210.35\pm 21.85$ & $0$ & $\infty$ & $0$\\ 
    \hline
         Q1 & Free & $99.3\pm 0.0$ & $2.0\pm 0.3$ & $2.85\pm 1.14$  & $0.31\pm 0.09$ & $\infty$ & $0$\\ 
         Q1 & XY4  & $99.3\pm 0.0$ & $52.4\pm 0.3$ & $11.15\pm 2.17$ & $0.13\pm 0.01$ & $\infty$ & $0$\\
         Q1 & CR-XY4 & $99.3\pm 0.0$ & $58.4\pm 0.9$ & $177.27\pm 40.70$ & $0$ & $\infty$ & $0$\\
         Q1 & XY4 + Idle & $99.4\pm 0.0$ & $51.8\pm 0.4$ & $18.24\pm 5.99$ & $0.40\pm 0.02$ & $\infty$ & $0$ \\
    \hline
         Q2 & Free & $99.4\pm 0.0$ & $53.7\pm 0.7$ & $6.90\pm 1.83$ & $0.09\pm 0.04$ & $\infty$ & $0$ \\ 
         Q2 & XY4  & $99.4\pm 0.0$ & $49.5\pm 0.6$ & $40.24\pm 1.45$ & $0.13\pm 0.00$ & $\infty$ & $0$ \\
         Q2 & CR-XY4 & $99.4\pm 0.0$ & $66.6\pm 0.9$ & $102.90\pm 10.11$ & $0$ & $\infty$ & $0$\\ 
         Q2 & XY4 + Idle & $99.4\pm 0.0$ & $67.8\pm 0.9$ & $104.99\pm 12.16$ & $0$ & $\infty$ & $0$\\ 
    \hline 
         Q3 & Free & $98.5\pm 0.0$ & $51.8\pm 0.1$ & $2.21\pm 0.83$ & $0.1\pm 0.37$ & $\infty$ & $0$\\
         Q3 & XY4  & $98.5\pm 0.0$ & $51.5\pm 0.2$ & $21.44\pm 3.53$ & $0.33\pm 0.01$ & $14.82\pm 1.77$ & $1.48\pm 0.26$ \\
         Q3 & CR-XY4 & $98.4\pm 0.0$ & $52.3\pm 0.4$ & $103.18\pm 18.93$ & $0$ & $\infty$ & $0$\\ 
         Q3 & XY4 + Idle & $98.5\pm 0.0$ & $55.2\pm 0.6$ & $979.49\pm 536.73$ & $0$ & $\infty$ & $0$\\
    \hline
         Q4 & Free & $99.0\pm 0.0$ & $55.7\pm 0.2$ & $5.60\pm 0.90$ & $41.61\pm 72.40$ & $\infty$ & $0$ \\ 
         Q4 & XY4  & $99.0\pm 0.0$ & $53.9\pm 0.3$ & $23.44\pm 1.02$ & $0.19\pm0.00$ & $\infty$ & $0$\\
         Q4 & CR-XY4 & $99.0\pm 0.0$ & $59.4\pm 0.3$ & $30.75\pm 0.62$ & $0$ & $\infty$ & $0$\\
         Q4 & XY4 + Idle & $99.0\pm 0.0$ & $55.9\pm 0.2$ & $16.98\pm 5.65$ & $0.23\pm 0.01$ & $7.68\pm1.42$ & $2.03\pm 1.18$ \\
    \hline\hline
    \end{tabular}
    \caption{Fit parameters when Eq. \ref{full_mod} is used to fit Fig. \ref{fig:sps-supp} in the main text. The decay times $\lambda$ significantly increase when implementing CR-XY4. The modulation frequency $\gamma$ vanishes for all CR-XY4 evolution as it mitigated quantum crosstalk that causes oscillation.}
    \label{tab:fit_param_supp}
\end{table*}

\subsection{Statistical Method} \label{bootstrapping}

The results reported in the main text display the mean and confidence intervals estimated via the bootstrapping method described in \cite{stine1989bootstrap}. This technique was implemented by randomly sampling $N$ data points (with replacement) from a data set of size $N$ and then computing the mean of this bootstrapped sample. By repeating this procedure $K$ times, a new, bootstrapped data set of size $K$ was generated. The mean and confidence interval (CI) can be calculated based on this bootstrapped data set. 

\begin{figure}[h]
    \centering
    \includegraphics[width=\columnwidth,]{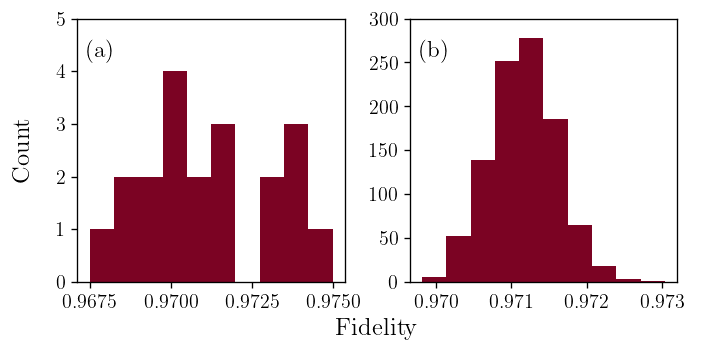}
    \caption{An example of the bootstrapping methodology. (a) 20 fidelity measurements of $K=1$ entangled states after $M=1$ \crxy4 sequences on all 4 Bell states with 5 repetitions each. (b) 1000 new data points after bootstrapping.}
    \label{fig:boots-ex}
\end{figure}

The same bootstrapping method is used for the data analysis of quantum state preservation experiments for both SPSs and MESs, and simultaneous QNS experiments. Using MESs preservation as an example: for the Bell state results shown in Fig. \ref{fig:mes-supp} (a)-(d), there were 20 initial data points for each value of $M$ (4 Bell states with 5 experiments each). The data was randomly sampled (with replacement) to generate a new data set with 20 data points before calculating the mean of this bootstrapped sample. This was repeated 1000 times to generate a bootstrapped data set. The mean and CI were then calculated from this bootstrapped data set. An example of bootstrapping is given in Fig.~\ref{fig:boots-ex}. 

\begin{figure*}[t!]
    \centering
    \includegraphics[width=\textwidth]{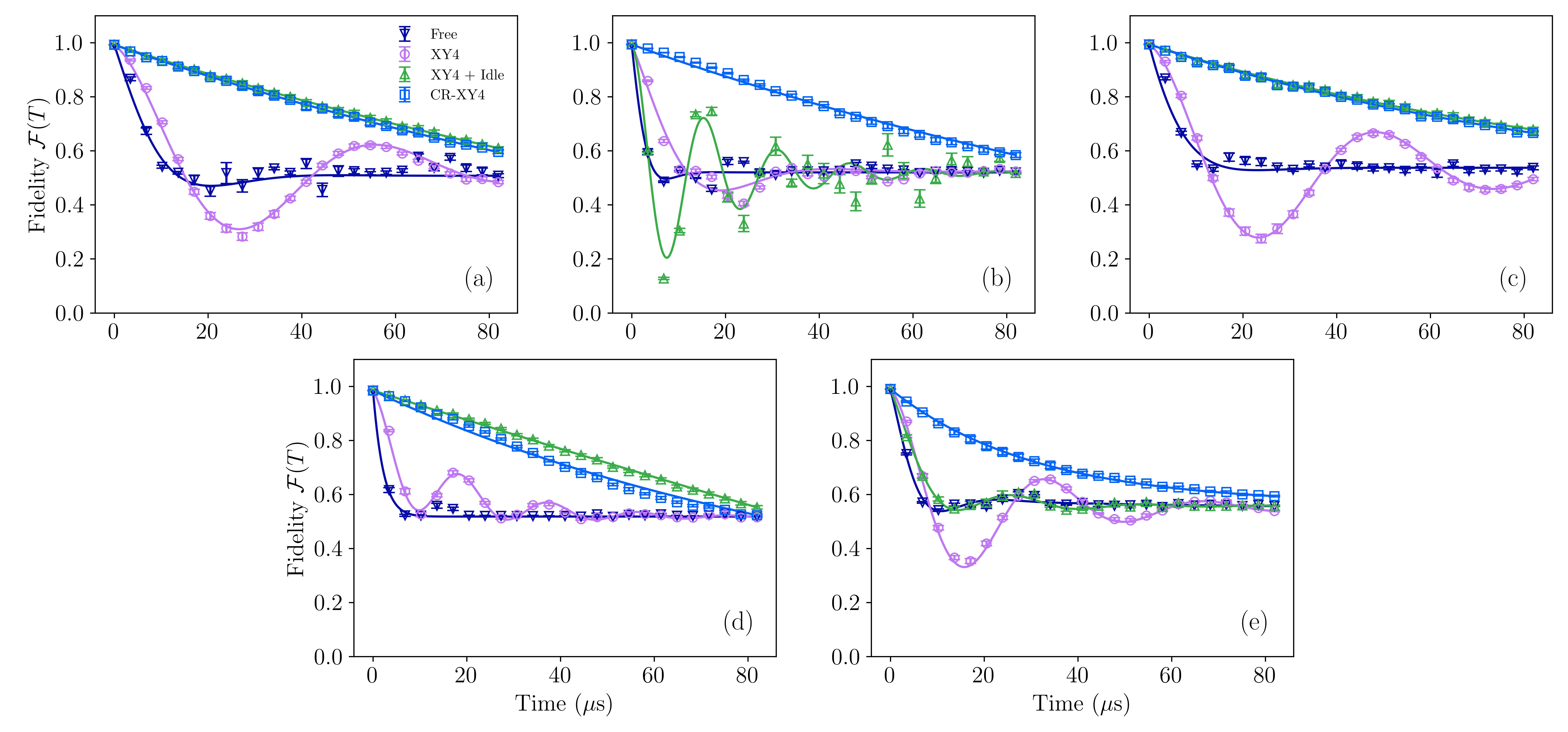}
    \caption{Simultaneous preservation of SPSs using different control protocols. Individual qubit fidelity vs. time for Q0 (a), Q1 (b), Q2 (c), Q3 (d), Q4 (e) of the IBMQE Lima 5-qubit processor using free evolution (dark blue down-triangles), XY4 (light purple circles), XY4 + Idle (light green triangles), and CR-XY4 (light blue squares). Data points and error bars denote mean fidelity and CIs, respectively, obtained from bootstrapping. }
    \label{fig:sps-supp}
\end{figure*}

\subsection{Additional Experimental Results}

\begin{figure*}[]
    \centering
    \includegraphics[width=\textwidth]{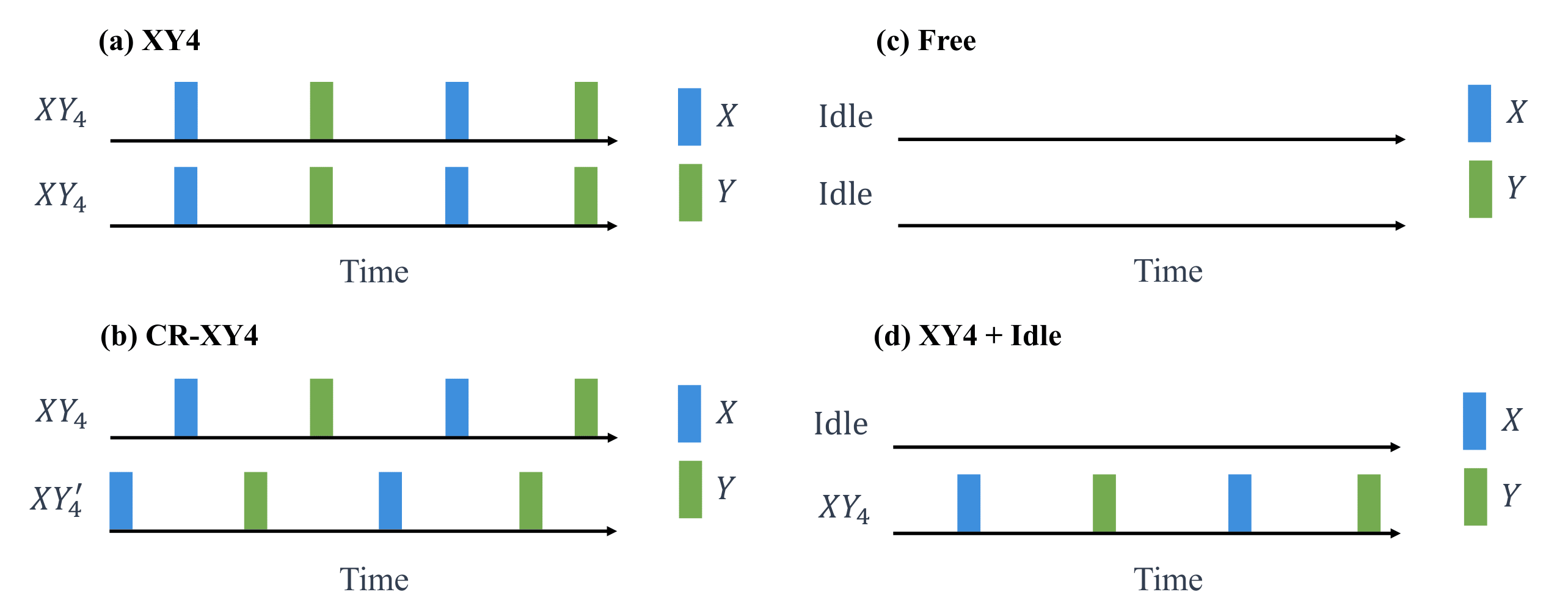}
    \caption{Pulse Schemes for four types of time evolution experimentally examined: XY4 (a), CR-XY4 (b), Free evolution (c), and XY4+Idle (d)}
    \label{fig:pulse_schemes}
\end{figure*}

\subsubsection{Single-Qubit Product States}
Using the IBMQE Lima processor, We experimentally investigate four types of time evolution including free evolution, XY4, CR-XY4, and XY4 + Idle (a crosstalk suppression sequence proposed in Ref.~\cite{Vinay2021DD}). The pulse schemes of the four types of time evolution are illustrated in Fig.\ref{fig:pulse_schemes}. XY4 + Idle takes the form of alternating XY4 and free evolution on neighboring qubits. According to the connectivity of quantum devices, we implement XY4 on Q0, Q2, and Q3, and free evolution on Q1 and Q4. Each experiment had 8000 shots and was repeated four times over the course of a four-day period from 3/29/22 to 4/01/2022. The individual qubit fidelity decay is shown in Fig. \ref{fig:sps-supp}.
Average fidelity and 95\% confidence intervals (CIs) are determined via the same bootstrapping procedure discussed above.

For individual qubit fidelity, CR-XY4 outperforms free evolution and XY4 in all cases. CR-XY4 outperforms XY4-Idle in fidelity decay and magnitude for Q1 and Q4. In contrast, both protocols maintain similar fidelity decay rates for Q0, Q2, and Q3. This finding can be easily derived analytically. Due to the asymmetric nature of the XY4 sequence, the switching function of the $i$th qubit $R_i(t)$ behaves as an odd function with respect to $T/2$, where $T$ is the total evolution time. And the switching function of its neighboring qubit remains unity throughout the evolution time as it is freely evolving; thus, $R_j(t)=1$. As a result, the overlap integral $\chi_{ij}(T) = J\int_0^TR_i(t)R_j(t)dt$ is odd over the interval $t\in [0,T]$ and therefore, zero.

The individual fidelity decay curves for the four types of time evolution are fit with the model Eq.~(\ref{full_mod}). Among twenty curves for five qubits, 18 curves can be well fit with the simplified model $f(t) = e^{-t/\lambda}\cos(\gamma t)$, with the other two requiring additional parameters to account for their dynamics. The fit parameters are reported in Table \ref{tab:fit_param_supp}. 
\subsubsection{Multipartite Entangled States}
\begin{figure*}[t]
    \centering
    \includegraphics[width=0.9\textwidth]{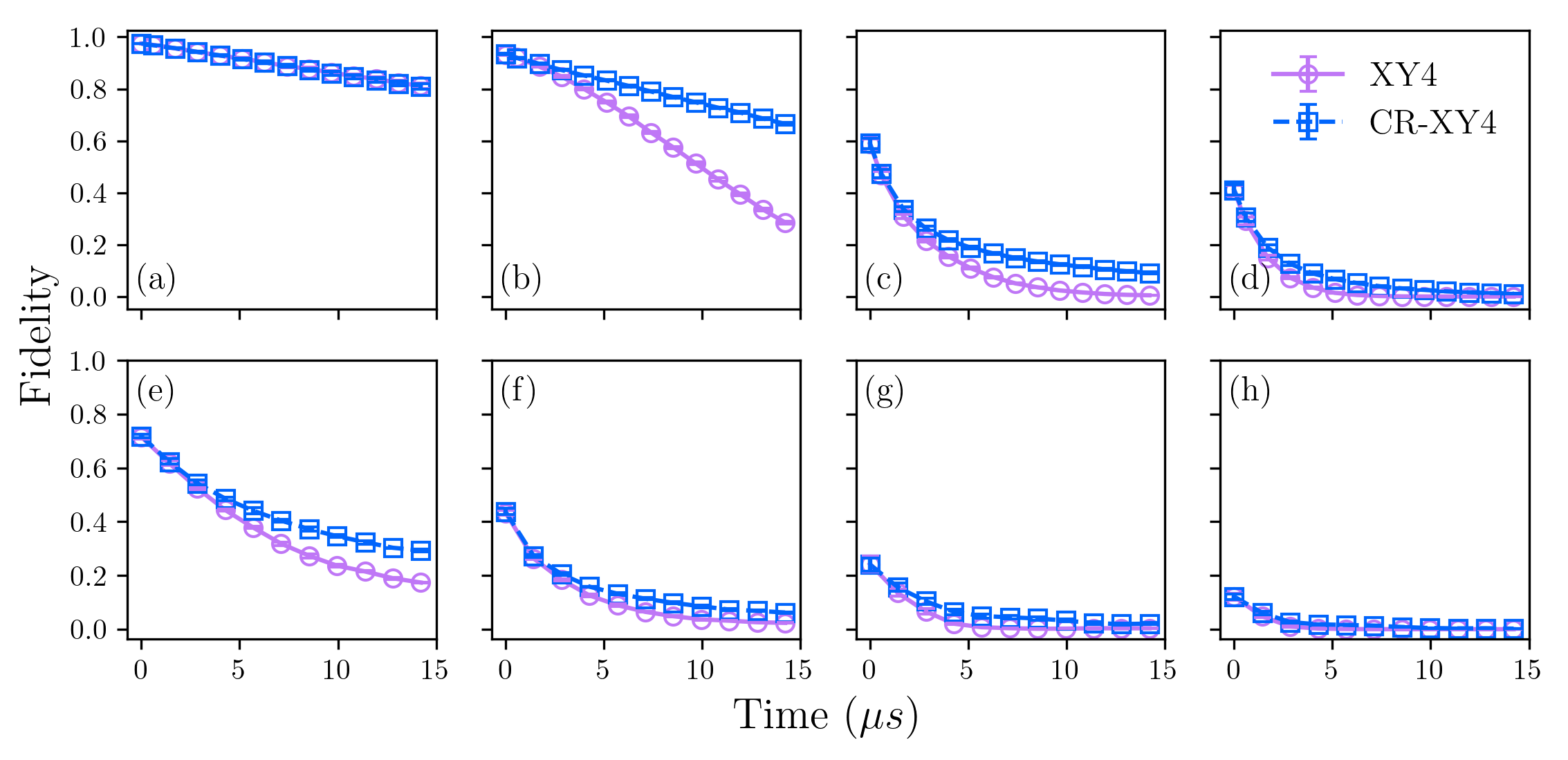}
    \caption{Sample of the Bell and $W$ state preservation experimental fidelity results performed on the IBM Auckland device. (a-d) Preservation of one, two, six, and ten Bell states respectively. (e-h) Preservation of one, two, four, and seven W states respectively. Each panel displays the bootstrapped results and 95\% confidence intervals from all experiments. There were 8000 shots per experiments and each experiment was repeated five times. Results indicate \crxy4 outperforms \xy4 in all cases except the single Bell state case.}
    \label{fig:mes-supp}
\end{figure*}
To demonstrate state preservation of multipartite entangled states (MES), we consider five different MESs including the two-qubit Bell states of the form: $\ket{\Phi_{\pm}}=1/\sqrt{2}(\ket{00}\pm\ket{11})$ and $\ket{\Psi_{\pm}}=1/\sqrt{2}(\ket{01}\pm\ket{10})$ and the three-qubit $W$ state $\ket{W}=1/\sqrt{3}(\ket{001}+\ket{010}+\ket{100})$. Each experiment consists of applying an increasing number of DD sequences in order to assess how long the sequences could preserve the state. In addition to increasing the number of DD sequences, we increased the number of MESs to assess how many states the DD sequence could preserve simultaneously. All MES preservation experiments were performed on the 27-qubit IBM Auckland device. Each experiment had 8000 shots and was repeated five times over the course of a five-day period starting on 4/29/22. This resulted in ten experiments for each initial state. The device topology is shown in Fig. \ref{fig:lima_nairobi}  and the qubit pairings for the entangled states are shown in Table \ref{tab:qubit-pairs}. 

\begin{table}[b]
    \centering
    \begin{tabular}{||p{3cm}|p{3.5cm}||}
    \hline
    \hline
    Bell States & $W$ State \\
    \hline
    \hline
         (1,4), (7,10), & (12,13,14), (15,18,17), \\
         (12,15), (18,21), & (21,23,24), (22,25,26), \\
         (23,24), (25,22), & (16,19,20), (9,8,11), \\
         (19,16), (14,11), & (2,3,5), (0,1,4), \\
         (8,5), (3,2) & (6,7,10) \\
    \hline
    \end{tabular}
    \caption{A list of the qubits paired for the Bell and $W$ states on the IBM Auckland device.}
    \label{tab:qubit-pairs}
\end{table}

The results from each experiment were used to produce fidelity versus time curves where the time is given by $T=M t_{\xy4}=M t_{\crxy4}$ and $M$ is the number of repetitions of the DD sequence. Applying the bootstrap method (\ref{bootstrapping}), we estimated the mean fidelity and 95\% confidence interval for the Bell and $W$ state results. A sample of these results are presented in Fig. \ref{fig:mes-supp}. As mentioned in the main text, a single Bell state (two qubits) is inherently invariant under the effects of $ZZ$-crosstalk which results in nearly identical performance between \xy4 and \crxy4 as shown in Fig. \ref{fig:mes-supp}(a). By placing two Bell states (four qubits) adjacent to each other on the quantum device, neighboring Bell pairs experience the effects of crosstalk across the common edge. Once the second Bell state is added, we observe that \crxy4 begins to outperform \xy4 after $M = 8$ applications of the DD sequences and continues to outperform as $M$ increases as shown in Fig.~\ref{fig:mes-supp}(b). As we increase the number of Bell states, $K$, we note that the initial fidelity decreases due to the increasing errors in state preparation. Despite this, we still observe that \crxy4 outperforms \xy4 in the case of $K=2,\ldots,10$.

Unlike the Bell states, the $W$ state is susceptible to the effects of $ZZ$-crosstalk. In the case of one W state (three qubits), we observe that \crxy4 outperforms after $M = 5$ sequences and continues as $M$ increases as shown in Fig. \ref{fig:mes-supp}(e). Again, we note that increasing the number of entangled states decreases the initial fidelity but \crxy4 continues to outperform \xy4 in all cases $K=1,\ldots,9$. 

\begin{figure}[t]
    \centering
    \includegraphics[width=\columnwidth]{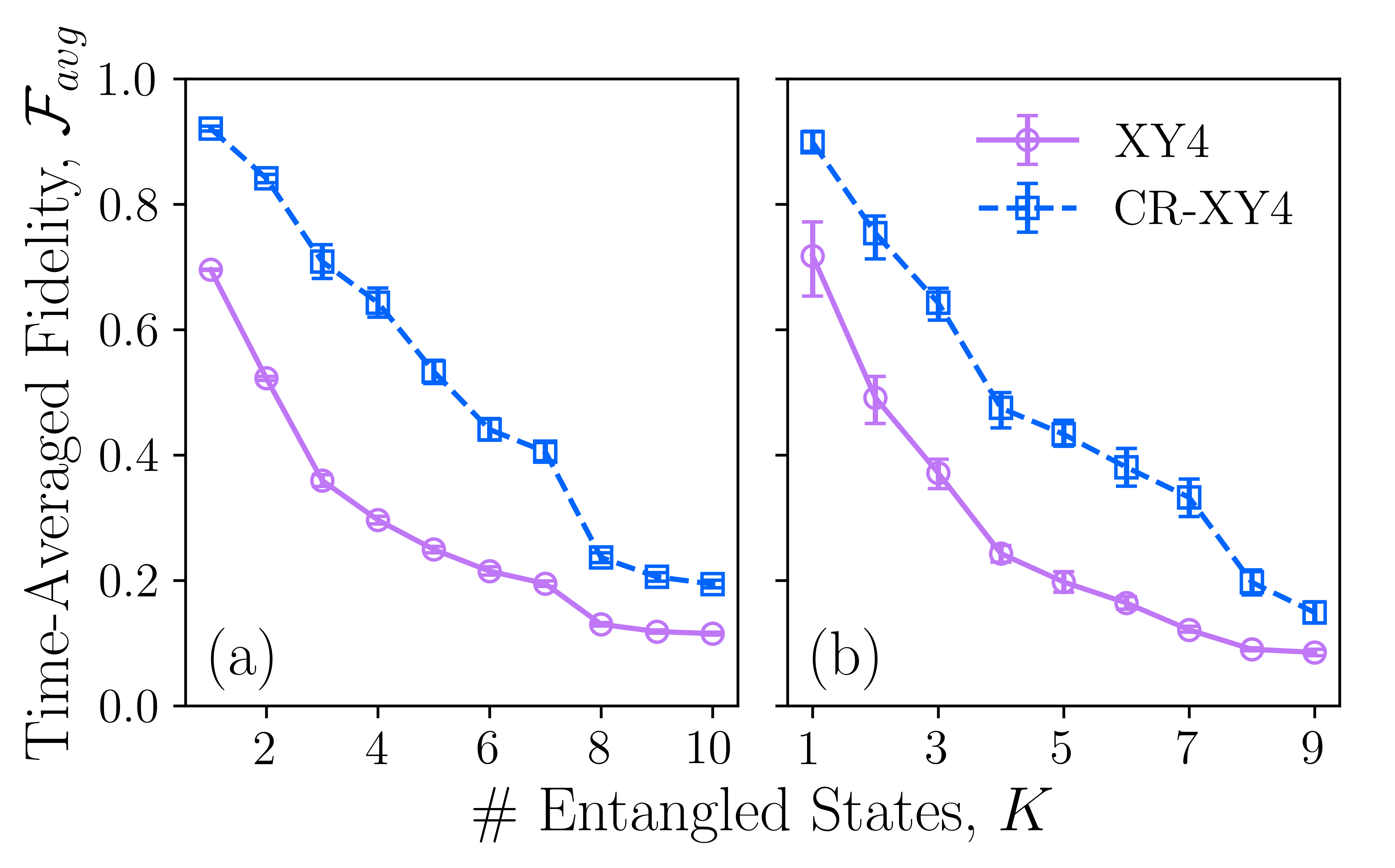}
    \caption{Time-averaged fidelity results for each value of $K\leq K_{max}$ for the Bell and $W$ state experiments. For the Bell states, $K_{max} = 10$ states on 20 qubits. For the W state, $K_{max} = 9$ states on 27 qubits. Each value of $\mathcal{F}_{\rm avg}$ corresponds to the best-performing chain of states for each $K$. Results indicate \crxy4 outperforms \xy4 for all values of $K$ for both MESs.}
    \label{fig:mes-scaling-sup}
\end{figure}

We further quantify the simultaneous preservation of the MESs using the time-averaged fidelity $\mathcal{F}_{\rm avg}$ given by $\mathcal{F}_{\rm avg}= T^{-1}_{\max}\int_0^{T_{\max}} \frac{\mathcal{F}(t)}{\mathcal{F}(0)}dt$. This is achieved by mining the experimental results for $K_{\textrm{max}}$ entangled states and identifying the best performing chain of states from $K = 1,\ldots,K_{\textrm{max}}$ where $K_{\textrm{max}}$ is 9 and 10 for the $W$ state and Bell states, respectively. We define a chain of states as a sequence of states that are prepared physically adjacent to each other on the quantum device as these are most likely to experience the effects of quantum crosstalk. As an example, for $K = 2$, we calculate $\mathcal{F}_{\rm avg}$ for every unique chain of two entangled states and chose the pair with the highest value. The bootstrapping method described in \ref{bootstrapping} is applied to $\mathcal{F}_{\rm avg}$ to determine the mean and CI for each $K$. By repeating this for all values of $K$ and for both DD sequences, we compare the performance of the \crxy4 and \xy4 as shown in Fig \ref{fig:mes-scaling-sup}. From these results, it is apparent that \crxy4 outperforms \xy4 for all values of $K$ for both the Bell and $W$ states. Using these results, we then calculate the fidelity ratio $R_\mathcal{F}=\mathcal{F}^{\crxy4}_{\rm avg}/\mathcal{F}^{\xy4}_{\rm avg}$, the results of which are shown in the main text.

\section{Quantum Noise Spectroscopy}
\subsection{Noise Injection}
To validate the efficacy of the QNS protocol and its crosstalk-robust variant, we used a Schrödinger Wave Autoregressive Moving Average (SchWARMA) model \cite{Schultz2021schwarma} to inject narrow band dephasing noise on each qubit individually, each having distinct center frequencies. Dephasing noise injection is implemented at the gate-level, where an $Rz$ gate is added after each control gate of the circuit. The noise injected circuit can be expressed as
\begin{eqnarray}
    U(\phi) = R_z(\phi_N)G_{N}R_z(\phi_{N-1})G_{N-1}...R_z(\phi_1)G_1, \quad
\end{eqnarray}
where $G_i$ represents noiseless single-qubit operations and $R_z(\phi_j)$ denotes a $z$ rotation with phase $\phi_j$~\cite{Schultz2021schwarma}. In our experiments, $G_i$ are chosen to denote the gates of the FTTPS protocol. A SchWARMA model is used to generate a trajectory of temporally correlated phases $\boldsymbol{\phi} = \{\phi_1, \phi_2, ..., \phi_N\}$ for each qubit noise process.

We perform noise characterization on $N=2$ to $7$ qubits using FTTPS and a CR variant. For each defined qubit noise process, 10 trajectories are generated and experimentally implemented. The state of the qubits are measured simultaneously at the end of the protocol. Data collected from the 10 realizations of the experiment are averaged to mimic the desired dephasing dynamics.

\subsection{Crosstalk Robust FTTPS}
Crosstalk Robust FTTPS (CR-FTTPS) enables simultaneous characterization of dephasing noise on multiple qubits with a built-in crosstalk suppression mechanism based on our analytical condition. Specifically, the control matrix of canonical FTTPS takes the form of $R_1(t) = \textrm{sign}(\cos(\omega_j t))$, with $\omega_j$ denoting the sweeping frequency of each sequence. Due to the property that cosine and sine functions are naturally even and odd functions, we design its CR counterpart to have a control matrix of $R_2(t) = \textrm{sign}(\sin(\omega_j t))$. In this way, the function quantifying the crosstalk $\chi(T) = J\int_0^T R_1(t)R_2(t)dt = 0$. The pulse sequences consists of identity gates for delay and $X$ gates implemented at each time $t$ to flip $R(t)$ between $+1$ and $-1$, according to their corresponding control matrices. The CR-FTTPS protocols is implemented on any neighboring qubits according to the topology of each device. In a seven-qubit experiment on IBMQ Nairobi, for example, the $R_1(t)$ sequence is implemented on Q0, Q2, Q3, Q4, Q6, while the $R_2(t)$ sequence is implemented on Q1 and Q5.

\subsection{Crosstalk Susceptibility and FTTPS}
\begin{figure}[t]
    \centering
    \includegraphics[width=\columnwidth]{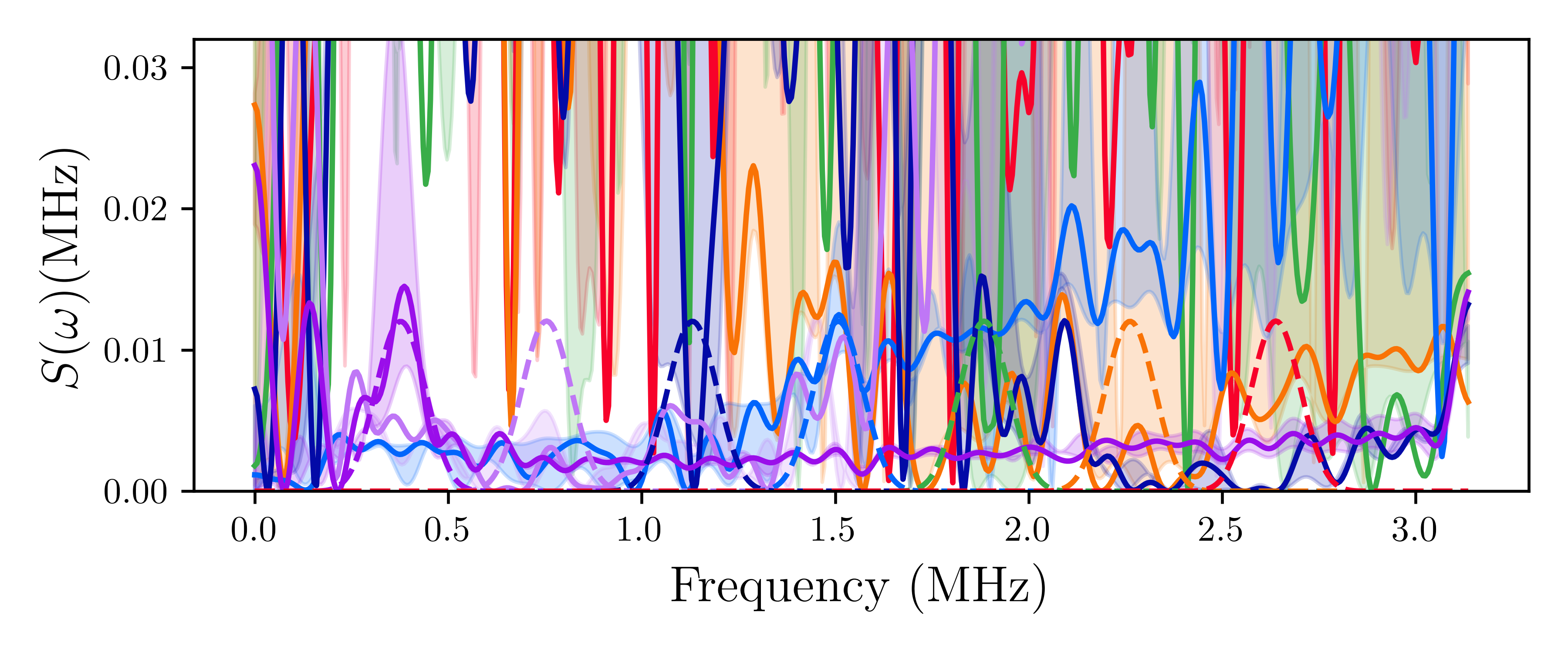}
    \includegraphics[width=\columnwidth]{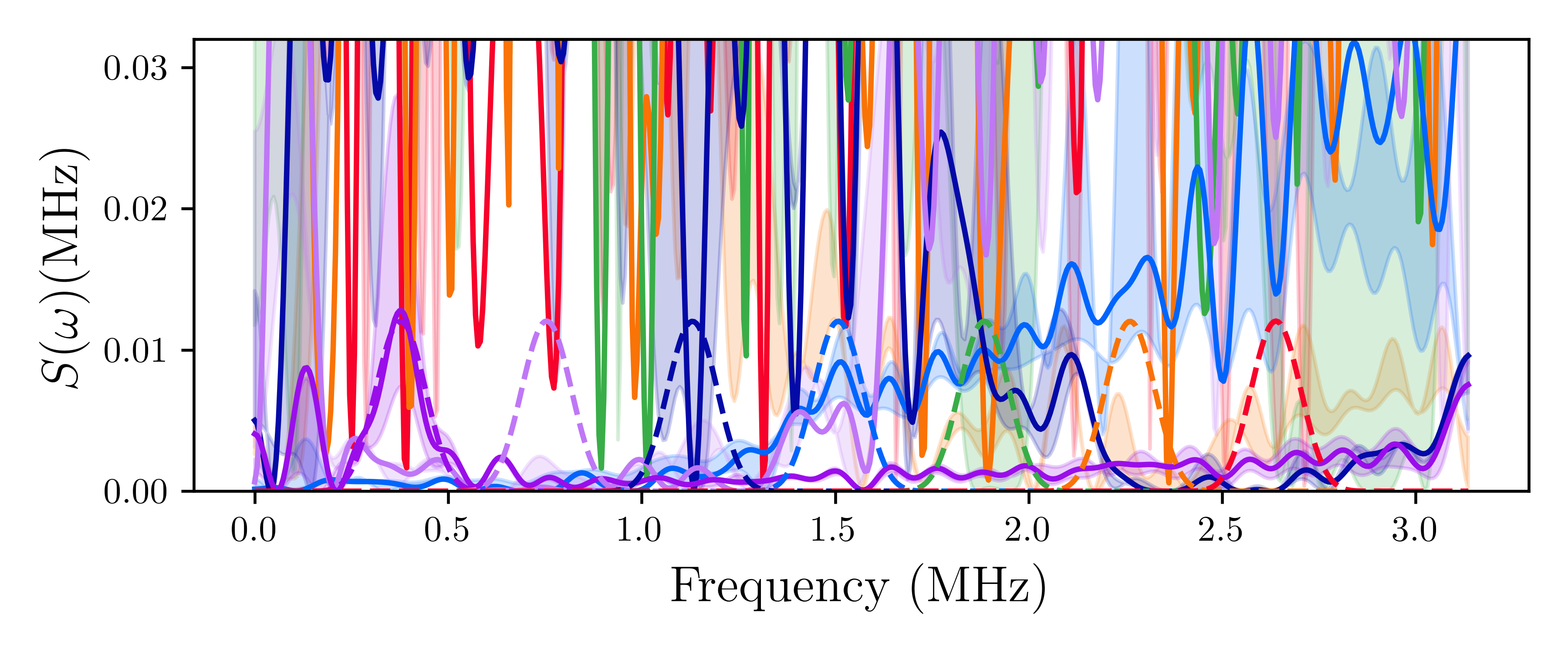}
    \caption{Canonical FTTPS (Top) and crosstalk-corrected FTTPS (Bottom) on the 7-qubit IBMQE Nairobi processor, where narrowband dephasing noise with distinct center frequencies is injected on each qubit. Average spectrum estimates
(solid lines) and CIs (shaded regions) indicate almost no agreement with
injected noise (dashed lines).}
    \label{fig:FTTPS_spec}
\end{figure}

When the canonical FTTPS protocol is applied to seven qubits simultaneously, static quantum crosstalk offsets the survival probabilities of each qubit and results in additional bit flips. As such, the spectrum estimates are subject to artificial increases in the noise floor and potentially significant spurious features. An example of the canonical FTTPS reconstruction is shown in Fig.~\ref{fig:FTTPS_spec} (Top) for a 7-qubit experiment.

While the canonical FTTPS protocol is susceptible to quantum crosstalk, it is possible to exploit our knowledge of the noise model to attempt a post-processing correction. The expectation value of $\sigma_x$ for the FTTPS protocol for qubit $i$ is determined by the first and second order cumulant according to:
$\langle \sigma_i^x \rangle = e^{-\mathcal{C}_O^{(2)}/2}\cos(2JT)$, 
where $J$ denotes the coupling strength between two connected qubits (retrieved from the IBMQ backend) and $T$ denotes total time of a circuit. Through measurement, we obtain survival probabilities $p^+$ for the equal superposition state for each qubit after executing each circuit. By the relationship $p^+ - p^- = \langle \sigma_i^x \rangle$ and $p^+ + p^- = 1$, we can relate the measured survival probabilities of the $\ket{\pm}$ states to the ``expected" decay model. Specifically, the model is used to adjust for the offset coefficient $\cos(JT)$ due to the crosstalk dynamics and therefore, estimate a crosstalk-corrected $p^+$. Spectral estimation is then performed with said crosstalk-corrected survival probabilities. An example reconstruction is shown in Fig.~\ref{fig:FTTPS_spec} (Bottom) for a 7-qubit experiment. Note that we observe little improvement from the model correction, indicating that the candidate model is likely not sufficient for modeling the dynamics. We hypothesize that additional noise generated by finite pulse width results in crosstalk along additional axes beyond $ZZ$. We explore this analytically for FTTPS and the CR variant in the next section.

\section{Analysis of Finite Pulse-Width Effects}

\begin{figure}[t]
    \centering
    \includegraphics[width=\columnwidth]{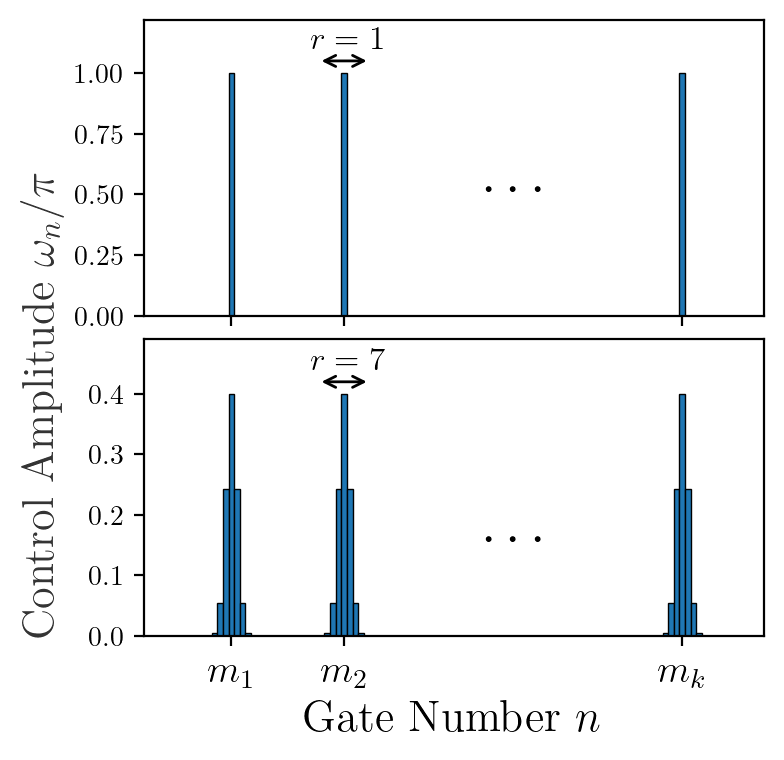}
    \caption{Schematics of relaxing the ideal pulse width condition, where pulse width is $r=1$, to $r=7$ with Gaussian pulse shape.}
    \label{fig:finite_width_pulses}
\end{figure}

In this section, we study the effects of relaxing the ideal pulse condition to show that the CR sequences provide crosstalk cancellation even in the presence of finite width pulses. 
Here, $N$ will denote the total number of gates in a sequence, where the gate number $n=0,...,N-1$ also represents the number of time steps elapsed. We denote by $\omega_n$ the control amplitude at the $n^{\mrm{th}}$ time step. 
In what follows, the index $m$ represents the center location of the pulses, which in the single gate-per-operation case coincides with the location of the gate. A schematic of this can be seen in the top panel of Fig.~\ref{fig:finite_width_pulses}, for Gaussian pulse shapes.
In the finite width case, each pulse possesses a finite duration $r>1$. For simplicity, only odd integers $r$ will be considered.
Note for both FTTPS and XY4 sequences, that the set of amplitudes $\omega_n$ must satisfy the conditions $\sum_{n\in \mathcal{M}_m} \omega_n = \pi$ for $\mathcal{M}_m=\{m-\floor{\frac{r}{2}},...,m+\floor{\frac{r}{2}}\}$, and $\omega_n=0$ for $n\notin \mM_m$. 

\subsection{Finite Pulse-Width in CR-FTTPS}

In the FTTPS case, we specify the pulse locations by $m_\ell^u$, with $u=\{\cos,\sin\}$ and $\ell\leq 2K-1$, where $K=N/2$ is the total number of FTTPS sequences. 
For the $k^{\mrm{th}}$-FTTPS, the discrete time locations of the $\ell^{\mathrm{th}}$ pulse of cosine- and sine-type sequences are defined respectively from sign changes of the functions $s^{\cos}_n=\mrm{sign}\left[ \cos(\pi n k/K) \right]$ and $s^{\sin}_n=\mrm{sign}\left[ \sin( \pi n k/K) \right]$. Then, we write the $\ell^{\mrm{th}}$ pulse location of cosine sequences as $m^{\cos}_\ell=\floor{(\ell+\frac{1}{2})\frac{K}{k}}$, and equivalently for the sine sequences $m^{\sin}_\ell=\floor{\ell\frac{K}{k}}$. 
Note that for a cosine pulse, $\ell=0,...,2k-1$, whereas for a sine pulse $\ell = 1,...,2k-1$. Therefore, we write the $k^{\mrm{th}}$-FTTPS cosine and sine controls as
\eq{
\Omega_{n}^{(k),u} = 
\begin{cases}
      \omega_{n} & \mrm{if} \quad |n-m^u_\ell|<\ffr \\
      0 & \mrm{otherwise}.
\end{cases} 
}
The noiseless time propagators can be computed as the time-ordered operator $U^{(k),u}_n = \prod_{p} e^{-\frac{i}{2} \Omega_p^{(k),u} \sigma_x}$. 
Following the formalism described in the main text, the single-qubit control matrices for each control type are defined as $R^{(k),u}_{\mu\nu,n} = \frac{1}{2} \mathrm{Tr}\left[ U^{(k),u}_n  \sigma_\mu U^{(k),u\dagger}_n \sigma_\nu \right]$.

In the single gate per pulse case, where $r=1$, the diagonal elements of the control matrices will alternate between $\pm 1$ at the location of each pulse. This can be seen by considering that the time propagators will alternate between identity operators $I$ and  $\sigma_x$ rotations at each pulse application $m^u_\ell$. Also note that the $X$ pulses can only act non-trivially along orthogonal axes. More specifically, when $\mu=x$, the operator $U_u^{(k)}(t)$ will commute with $\sigma_x$, yielding $R_{x\nu}^{(k),u}=\frac{1}{2} \mathrm{Tr}\left[ \sigma_x \sigma_\nu \right] = \delta_{x\nu}$, where the analogous holds for $\nu=x$. Therefore, in what follows we will focus on $\mu,\nu=y,z$. Noting that $\sigma_x \sigma_\mu \sigma_x = - \sigma_\mu$ for $\mu,\nu=y,z$, the control matrices become
\eq{
\label{eq:Rmn_ideal}
R^{(k),u}_{\mu\nu,n} = 
\begin{cases}
    \delta_{\mu\nu}   & \mrm{if} \quad U^{(k),u}_n \doteq I \\
    -\delta_{\mu\nu} & \mrm{if} \quad U^{(k),u}_n \doteq \sigma_x,
\end{cases}
}
where the symbol $\doteq$ denotes equality between operators up to a global phase.

As shown below, when the ideal pulse condition is relaxed, the control matrices during the applications of pulses become discrete interpolations between these values of $\pm\delta_{\mu\nu}$.
First, let us look at an odd $\ell$ pulse case, where the accumulated ideal time evolution $U_n^{(k),u}$ performs an identity operation.
During the application of the pulse, the control matrices can be written as
\eq{
\label{eq:Rmn_interp}
R^{(k),u}_{\mu\nu,n} &\doteq
\begin{pmatrix}
\cos\alpha^{m}_{n} & \sin\alpha^{m}_{n}\\
-\sin\alpha^{m}_{n} & \cos\alpha^{m}_{n}
\end{pmatrix} \\
&= R_X(\alpha_{n}^{m}),
}
where $\alpha_{n}^{m}=\sum_{p=m-\ffr}^{m+\ffr} \omega_n$. 
Note that $R^{(k),u}_{\mu\nu,m-\ffr-1}=\delta_{\mu\nu}$, whereas $R^{(k),u}_{\mu\nu,m+\ffr+1}=-\delta_{\mu\nu}$. This implies that $R^{(k),u}_{\mu\nu,n}$ is a time-dependent rotation matrix interpolating between $+ \delta_{\mu,\nu}$ and $- \delta_{\mu,\nu}$. 
For an even $\ell$ pulse, it is easy to see that $R^{(k),u}_{\mu\nu,n}=-R_X(\alpha_n^m)$, hence taking $-\delta_{\mu\nu}$ to $+\delta_{\mu\nu}$.

\begin{figure*}[]
    \centering
    \includegraphics[width=2\columnwidth]{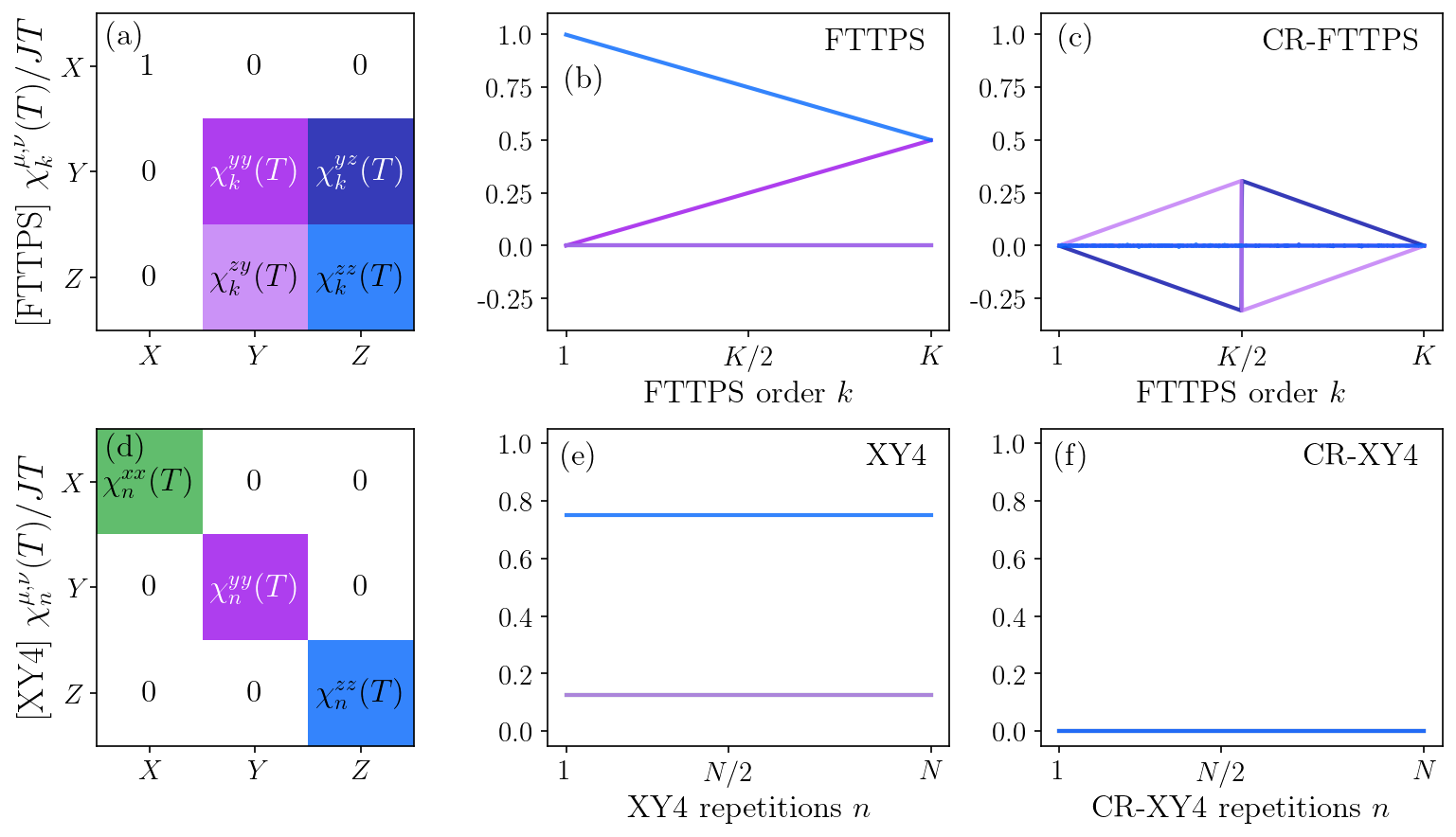}
    \caption{Numerical results for $\chi^{\mu\nu}(T)$ obtained from relaxing the ideal pulse condition on FTTPS sequences (top) and XY4 (bottom), with $\delta t = 5 T/N$. Square pulses were assumed. The left panels (a) and (d) show the respective non-trivial overlap components. Both simultaneous FTTPS/XY4 (see panels (b) and (c)) and CR-FTTPS/CR-XY4 (see panels (e) and (f)) are shown. The simulation is implemented utilizing a gate-based structure, which provides time discretization. Due to numerical resolution constraints in the pulse times $t_m^u$, for FTTPS sequences with  $k\in[K/2,K]$ the sequences possess an odd number of $X$ pulses. This leads to the discrete jump observed at $k=K/2$ for CR-FTTPS. Here, the maximum FTTPS order is $K=1024$, whereas the maximum number of XY4 repetitions is $N=100$.}
    \label{fig:finite_width}
\end{figure*}

Let us now look at the crosstalk condition Eq.~(\ref{eq:first_order_suppression_criteria}), and the effect finite pulse width has on both simultaneous FTTPS and CR-FTTPS sequences. Simultaneous FTTPS sequences are characterized by the implementation of $\Omega_{n}^{(k),\cos}$ on both qubits. In what follows, we will assume for simplicity that pulses are symmetric with respect to the center $m$, i.e., $\omega_{m-n}=\omega_{m+n}, n\leq\ffr$. Consequently, for the $k^{\mathrm{th}}$-FTTPS, the diagonal elements become
\eq{
\frac{\chi^{(k)}_{zz}(T)}{JT} &= \frac{1}{N} \sum_{n=0}^{N-1} R^{(k),\cos}_{zz,n} R^{(k),\cos}_{zz,n} = \nonumber\\
&\overset{(1)}{=} \frac{1}{N} \sum_{\ell=0}^{2k-1} \left( m^{\cos}_{\ell+1} - m^{\cos}_\ell \right) + \nonumber\\ 
&\quad + \sum_{n=m^{\cos}_\ell-\ffr}^{m^{\cos}_\ell+\ffr} \cos^2 \left[ \sum_{p=m^{\cos}_\ell-\ffr}^{n}  \Omega^{(k),\cos}_p \right] = \nonumber\\
&= \left(1 - k \frac{r}{N}\right) + O(k/N) + \nonumber\\ 
&\quad + \frac{2k}{N} \sum_{n=-\ffr}^{\ffr} \cos^2 \left[ \sum_{p=-\ffr}^{n} \omega_p \right] = \nonumber\\
&= 1 - k \frac{(r-1)}{N} + O(k/N),}
\eq{
\frac{\chi^{(k)}_{yy}(T)}{JT} &= \frac{1}{N} \sum_{n=0}^{N-1} R^{(k),\cos}_{zy,n} R^{(k),\cos}_{zy,n} = \nonumber\\
&= \frac{k}{N} \sum_{n=-\ffr}^{\ffr} \sin^2 \left[ \sum_{p=-\ffr}^{n} \omega_p \right] + O(k/N) = \nonumber\\
&= k \frac{r}{N} + O(k/N).
}
We used that the free evolution contributions $m^{\cos}_{\ell+1} - m^{\cos}_\ell = K/k - r + O(1)$ in (1), where the $O(1)$ difference comes from evaluating the integer part. The trigonometric sums can be resolved approximately using the symmetric pulse assumption. 
Note that in the single gate pulse limit $r=1$, the values $\chi_{zz}(T)\approx JT, \chi_{yy}(T)\approx 0$ are recovered. 
The off-diagonal elements, on the other hand, remain close to zero even in the presence of finite width pulses
\eq{
\frac{\chi^{(k)}_{yz}(T)}{JT} &= \frac{1}{N} \sum_{n=0}^{N-1} R^{(k),\cos}_{zy,n} R^{(k),\cos}_{zz,n} = \nonumber\\
&= \frac{2k}{N} \sum_{n=-\ffr}^{\ffr} \cos \left[ \sum_{p=-\ffr}^{n} \omega_p \right] \times \nonumber\\
&\quad \times\sin \left[ \sum_{p=-\ffr}^{n} \omega_p \right] +O(k/N) = \nonumber\\
&= O(k/N),
}
where the difference from zero comes again from taking the integer part in the pulse locations. Hence, we observe that the effect of finite pulse-width on simultaneous FTTPS is to shift complexity from the $zz$ to $yy$ component of $\chi^{(k)}_{\mu\nu}(T)$. An example of this is shown in panel (b) of Fig.~\ref{fig:finite_width}, where it can be seen that the contributions $O(k/N)$ are in fact small.

Let us now switch focus to the CR-FTTPS protocol, where $\Omega_{n}^{(k),\cos}, \Omega_{n}^{(k),\sin}$ are implemented on the first and second qubit, respectively. As in the simultaneous FTTPS case, we can perform the analysis of the overlap coefficients, wherein the ideal pulse limit it is easy to see that $\chi^{\mu\nu}(T)=0$. Performing the analogous calculation for a $k^{\mathrm{th}}$-CR-FTTPS sequence yields,
\eq{
\frac{\chi^{(k)}_{zz}(T)}{JT} &=  \frac{1}{N} \sum_{n=0}^{N-1} R^{(k),\cos}_{zz,n} R^{(k),\sin}_{zz,n} = \nonumber\\
&= \frac{1}{N} \sum_{n=-\ffr}^{\ffr} \cos \left[ \sum_{p=-\ffr}^{n} \omega_p \right] + O(k/N), \\
\frac{\chi^{(k)}_{yz}(T)}{JT} &= \frac{1}{N} \sum_{n=0}^{N-1} R^{(k),\cos}_{zy,n} R^{(k),\sin}_{zz,n}  = \nonumber\\
&= \frac{2k}{N} \sum_{n=-\ffr}^{\ffr} \sin \left[ \sum_{p=-\ffr}^{n} \omega_p \right] + O(k/N), \\
\frac{\chi^{(k)}_{zy}(T)}{JT} &= \frac{1}{N} \sum_{n=0}^{N-1} R^{(k),\cos}_{zz,n} R^{(k),\sin}_{zy,n} = \nonumber\\
&=- \frac{(2k-1)}{N} \sum_{n=-\ffr}^{\ffr} \sin \left[ \sum_{p=-\ffr}^{n} \omega_p \right] + O(k/N),
}
\eq{
\frac{\chi^{(k)}_{yy}(T)}{JT} &= \frac{1}{N} \sum_{n=0}^{N-1} R^{(k),\cos}_{zy,n} R^{(k),\sin}_{zy,n} = \nonumber\\
&= O(k/N).
}
The explicit values of the matrix will depend on the specifics of the pulse shape $\omega_n$ and are in general hard to compute analytically due to the specific contribution of $O(k/N)$. Assuming a constant pulse $\omega_n=\pi/r$, we can see in the limit of $k \ll N$ that for the $k^{\mrm{th}}$ CR-FTTPS sequence
\eq{
\chi^{(k)}_{\mu\nu}(T) &\approx J T \frac{4kr}{\pi N} 
\begin{pmatrix}
0 & -1\\
1 & 0
\end{pmatrix}
}
with $\mu,\nu=y,z$. This implies that the diagonal elements remain zero even in the presence of finite pulse-width, whereas the off-diagonal elements grow linearly with $k$ (in absolute value). Away from the ideal pulse approximation, we conclude that implementing CR-FTTPS sequences instead of simultaneous FTTPS has the effect of shifting the complexity from the diagonal to the off-diagonal elements of $\chi^{(k)}_{\mu\nu}(T)$. 
Panel (c) of Fig.~\ref{fig:finite_width} summarizes these results in numerical simulation. Interestingly, we observe that for $k>K/2$ the contribution $O(k/N)$ becomes significant and reduces the effect of crosstalk.

\subsection{Finite Pulse-Width in CR-XY4}
Here, we analyze the effect of finite pulse-width in the XY4 protocols.
We choose a pulse-width duration matching the experiments presented in the main text, where all $X$ and $Y$ pulses have the same duration as the idle periods.
We denote the duration of an XY4 block by $\tau$, and the pulse duration by $\delta t$. Note that in the single gate-per-pulse limit, $\tau = 8 \delta t$. Since the XY4 protocol is characterized by $N$ repetitions of the sequence of operators $IXIYIXIY$, we write the finite-width control as
\eq{
\vec{\Omega}_n\cdot \vec{\sigma} = 
\begin{cases}
    0 & \mrm{if} \quad 1\leq n\leq r \\
    \frac{\pi}{r} \sigma_x & \mrm{if} \quad r+1\leq n\leq 2r \\
    0 & \mrm{if} \quad 2r+1\leq n\leq 3r \\
    \frac{\pi}{r} \sigma_y & \mrm{if} \quad 3r+1\leq n\leq 4r = \tau /2\delta t  \\
    \dots\\
   \frac{\pi}{r} \sigma_y & \mrm{if} \quad 7rN\leq n\leq 8rN = T/\delta t,
\end{cases} 
}
where $r$ is an integer representing the width of each pulse. The control repeats with time period $\tau/2$ for a total of $2N$ repetitions.
It is straightforward to check that after an iteration of the fundamental block control sequence ($1\leq n\leq 8r$) the operation performed is an identity gate, i.e., $U(\tau)=I$. Note that the total number of time steps (gates) is $T=8rN\delta t$.

Next, we show that it is possible to use the control periodicity to derive a comb-like structure of the control matrix, simplifying its analysis. When this periodicity to the identity is present, we can write a control time propagator as $U(t) = U(t-p\tau) U(p\tau) = U(t-p\tau)$, where $p=\lfloor t/\tau \rfloor$ such that $t-p\tau<\tau$. It is straightforward to see that this implies that the control matrix is also periodic in time $R_{\mu\nu}(t)=R_{\mu\nu}(t-p\tau)$. These results are easily extended to discrete (gate) time. We can then derive the following expression for the overlap
\eq{
\frac{\chi_{\mu\nu}(T)}{JT} &= \frac{1}{8rN} \sum_{n=1}^{8rN} R^{}_{z\mu,n} R^{}_{z\nu,n}  = \nonumber \\
&= \frac{1}{8rN} \sum_{p=0}^{N-1}
 \sum_{n=8pr}^{8(p+1)r} R^{}_{z\mu,n} R^{}_{z\nu,n} = \nonumber \\
&= \frac{1}{8rN} \sum_{p=0}^{N-1}
 \sum_{n=1}^{8r} R^{}_{z\mu,n} R^{}_{z\nu,n} = \nonumber \\
&= N \frac{\chi_{\mu\nu}(\tau)}{J\tau},
}
where we used explicitly that $\tau=8r(\delta t/8)$. This means that the overlap normalized by the sequence duration grows linearly with the number of repetitions. Hence, it is sufficient to analytically compute the overlap for the first block.

Borrowing from intuition described in the previous section, the control matrix can be computed in a straightforward manner. More explicitly, the application of each $X$($Y$) pulse implements a $\pi$-rotation of the control matrix around the $X$($Y$) axis. This results in a control matrix consisting of alternating $\pm 1$ values for even $m$ intervals $m r<n<(m+1)r$, and composition of rotating matrices with time-dependent angles $\alpha_{m,n}=\sum_{p=mr}^{n+mr-1} \pi/r = \pi n/r$, for $n<r$ and odd $m$ values. Note that the analogous occurs for an {\xyp} sequence, where the rotating angles occur for even values of $m$. Namely, for $n<8r$, and suppressing the gate number $n$ dependence on $\alpha_{m,n}$ for simplicity of notation,
\eq{
R_{\mu\nu,n} = 
\begin{cases}
    \begin{pmatrix}
    1&0&0\\
    0&1&0\\
    0&0&1\\
    \end{pmatrix}& \mrm{if} \quad 1\leq n\leq r  \\
    \begin{pmatrix}
    1&0&0\\
    0&\cos\alpha_1 &\sin\alpha_1\\
    0&-\sin\alpha_1 &\cos\alpha_1\\
    \end{pmatrix}& \mrm{if} \quad r+1\leq n\leq 2r  \\
    &\cdots \\
    \begin{pmatrix}
    -\cos\alpha_7&0&\sin\alpha_7\\
    0&1 &0\\
    -\sin\alpha_7& 0 &-\cos\alpha_7\\
    \end{pmatrix} & \mrm{if} \quad 7r+1\leq n\leq \tau.
\end{cases} 
}

The analogous computation can be performed for the {\xyp} sequence, where it is easy to see that it matches the \xy4 calculation shifted by $r$, i.e., $R_{\mu\nu,n}^{XY4'} = R_{\mu\nu,n-r}^{XY4}$. Then, the overlap for simultaneous XY4 can be computed utilizing a counting argument, from which we obtain that
\eq{
\chi_{\mu\nu}^{XY4}(T) = \frac{JT}{8} \, \mathrm{diag}(1,1,6),
}
where $T=N\tau$ was used. This is shown in numerical simulation in panel (e) of Fig.~\ref{fig:finite_width}.

Lastly, using an analogous argument for CR-XY4 it can be seen that
\eq{
\chi_{\mu\nu}^{CR-XY4}(T) = 0,
}
for all $\mu,\nu=x,y,z$, where the cancelations arise from vanishing integrals of cosines and summing integrals of sines with alternating signs. This result implies that CR-XY4 is robust to finite pulse-widths, and is confirmed through numerical simulation, shown in panel (f) of Fig.~\ref{fig:finite_width}.
\\


%
%
%

\end{document}